\documentclass[twocolumn,astrosymb,tighten,twocolappendix,longbib,trackchanges]{aastex7} 
\usepackage{mathrsfs}

\usepackage{nccmath,gensymb} 
\usepackage{xspace}

\newcommand{\apDIB}{15272~\r{A} DIB\xspace}
\newcommand{\gDIB}{8623~\r{A} DIB\xspace}
\newcommand\CaII{\ion{Ca}{2}\xspace}

\shorttitle{DIBs Beyond $R(V)$}
\shortauthors{Saydjari \& Green}

\graphicspath{{./}{figures/}}

\begin{document}

\title{Correlations between Dust Extinction Features across All Wavelength Scales: \\ From Diffuse Interstellar Bands to $R(V)$}

\author[orcid=0000-0002-6561-9002,gname=Andrew,sname=Saydjari]{Andrew~K.~Saydjari}
\altaffiliation{Hubble Fellow}
\affiliation{Department of Astrophysical Sciences, Princeton University,
Princeton, NJ 08544 USA}
\email[show]{aksaydjari@gmail.com}  
\correspondingauthor{Andrew~K.~Saydjari}

\author[0000-0001-5417-2260,gname=Gregory,sname=Green]{Gregory M. Green}
\affiliation{Max Planck Institute for Astronomy, K\"{o}nigstuhl 17, D-69117 Heidelberg, Germany}
\email{gregorymgreen@gmail.com}

\begin{abstract}
Understanding variations in the dust extinction curve is imperative for using dust as a tracer of local structure in the interstellar medium, understanding dust chemistry, and observational color corrections where dust is a nuisance parameter. However, the extinction curve is complicated and exhibits features across a wide range of wavelength scales, from narrow atomic lines and diffuse interstellar bands (``DIBs''), to intermediate-scale and very broad structures (``ISS'' and ``VBS''), and the overall slope of the optical extinction curve, parameterized by $R(V)$. Robust, population-level studies of variations in these features are only now possible with large, all-sky, spectroscopic surveys. However, these features are often studied independently because they require drastically different spectral resolution. In this work, we couple features with disparate wavelength scales by cross-matching precision catalogs of DIB measurements from APOGEE and Gaia~RVS with low-resolution extinction-curve measurements from Gaia~XP. Using this combination, we show that there are meaningful correlations between the strengths of extinction-curve features across all wavelength scales. We present a model that statistically explains part of the excess scatter in DIB strength versus extinction, and we show variation in line shapes of two DIBs as a function of $R(V)$. We find that most DIBs increase in strength with increasing $R(V)$ and/or increasing strength of the ISS, though we found one DIB that anomalously decreases in strength with increasing $R(V)$. Using the behavior of the ensemble of DIBs in APOGEE, we present this as the first evidence of systematic chemical variation accompanying $R(V)$ variation.
\end{abstract}


\section{Introduction} \label{sec:intro}

Interstellar dust absorbs and scatters background light in a wavelength-dependent manner. The ``extinction curve'' -- extinction as a function of wavelength -- varies throughout space, and exhibits features at a wide range of wavelength scales.\footnote{The extinction curve is sometimes also called a ``reddening law,'' but this may suggest that it is a power law in wavelength, or some universal law of nature. Both are far from true.} At the broadest wavelength scales, the overall slope of the optical extinction curve, typically parameterized by $R(V) \equiv A(V)/E(B-V)$, is observed to vary throughout the Milky Way \citep{Rieke_1985_ApJ, Cardelli_1989_Extinction_Curve_CCM89,Mathis_1990_ARA_A, Pei_1992_ApJ, Fitzpatrick_1999_interstellar_extinction, Indebetouw_2005_ApJ, Fitzpatrick_2019_extinction_curves, Gordon_2023_ApJ, Schlafly_2016_Rv_catalog, Schlafly_2017_Rv_3D_mapping, Zhang_2025_Sci}. At the narrowest wavelength scales, individual atomic and molecular species imprint narrow absorption features, such as sodium, calcium and potassium absorption lines \citep{Hobbs_1974_ISM_Na_Ca_K_absorption}. 

\begin{figure*}[htb!]
    \centering
    \includegraphics[width=\linewidth]{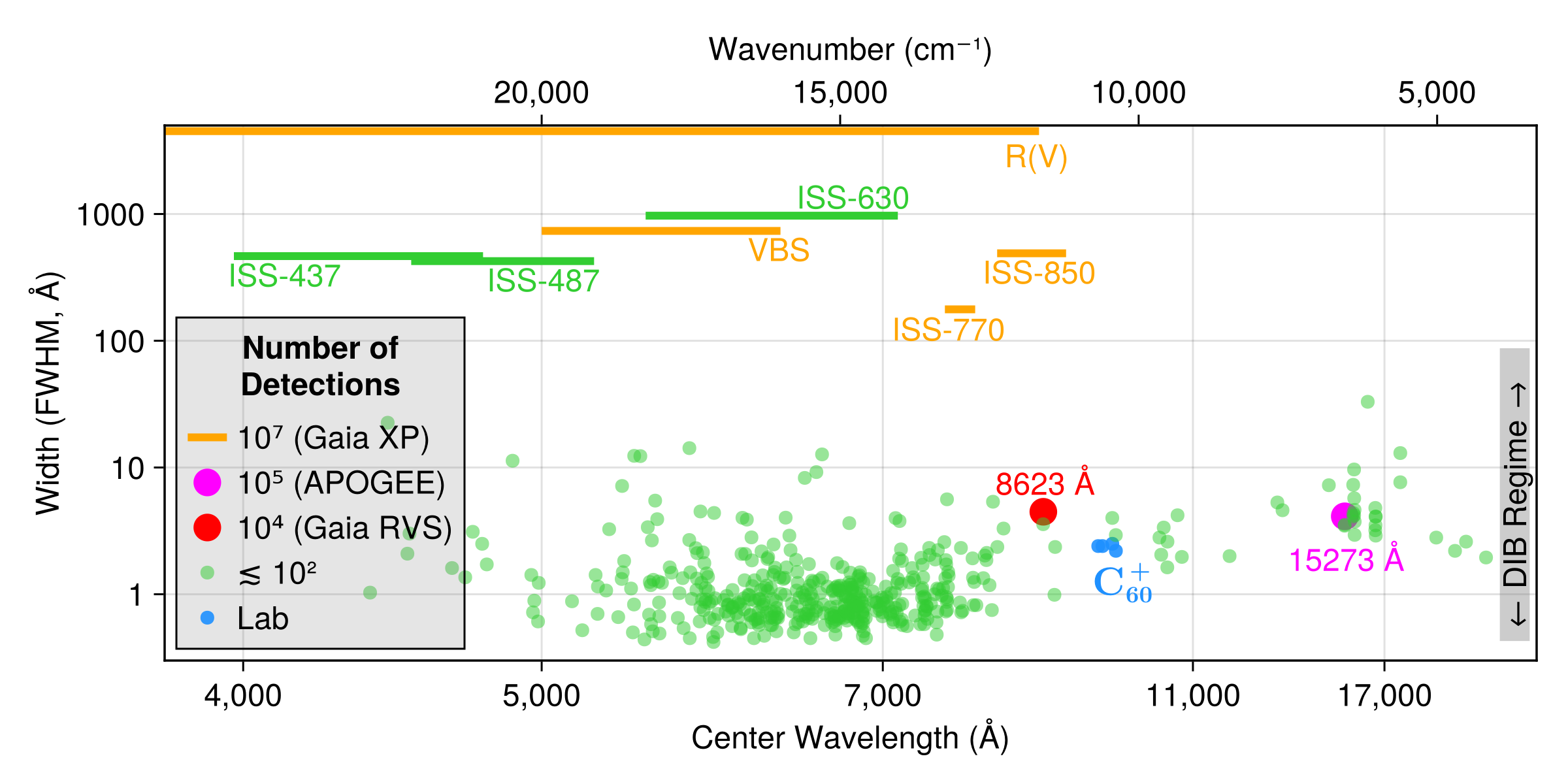}
    \caption{Overview of the size of catalogs of features in the optical-NIR extinction curve as a function of their width in wavelength, the widest being $R(V)$ which describes the slope of the optical extinction curve and the narrowest being diffuse interstellar bands (DIBs). The largest DIB catalogs are the \apDIB from APOGEE \citep[145,713 ][]{Saydjari_APOGEE_DIB_unpub} and \gDIB in Gaia RVS \citep[7,789][]{Saydjari_2023_ApJ}. We also highlight the 4 DIBs that have been confidently assigned to $C_{60}^+$ by laboratory measurements. While many other DIBs are known, most have only been measured in catalogs comprising a few stars. The largest catalog of wider wavelength features including intermediate scale structure (ISS), very broad structure (VBS), and $R(V)$ are from Gaia XP spectra \citep[130 million][]{Green_2024_arXiv}.}
    \label{fig:extinction_feature_overview}
\end{figure*}

Between these scales, there are a wide range of absorption features. Hundreds of diffuse interstellar bands (``DIBs''), with widths 0.4 -- 30\,\r{A}, have been detected \citep{McCall_2013_DIB_history}, though only one carrier ($\mathrm{C}_{60}^+$, known to be responsible for 4 -- 5 identified DIBs) has been conclusively identified \citep{Foing_1994_Natur, Campbell_2015_Nature_C60plus_DIB, Walker_2015_ApJL, lykhin2018electronic}. Several intermediate-scale structures (``ISS''), with widths 150 -- 500\,\r{A}, have been detected \citep{Massa_2020_optical_extinction_structure, MaizApellaniz_2021_7700AA_feature, Gordon_2023_ApJ}, though all are of unknown origin. The ``very broad structure'' (``VBS''), a depression in the extinction curve that roughly spans the range 5000 -- 6250\,\r{A}, is likewise of unknown origin \citep{Massa_2020_optical_extinction_structure}. In the ultraviolet, the 2175\,\r{A} feature, with a width of $\sim 500$\,\r{A}, is likely of carbonaceous origin (polycyclic aromatic hydrocarbons or graphite). These optical and near-infrared extinction curve features are summarized in Figure~\ref{fig:extinction_feature_overview}, using several DIB catalogs \citep{Hobbs_2008_ApJ, Cox_2014_AA, Elyajouri_2017_AA, Galazutdinov_2017_MNRAS}.

Correlations between the strengths of these various features are one tool for determining their origins. A high correlation between the strengths of two features could indicate that they share a common chemical carrier, or that their carriers form in similar environments. Searches have been conducted for correlations between equivalent widths (e.g., \citealt{Friedman_2011_DIB_correlations}) and shapes (e.g., \citealt{Ebenbichler_2024_EDIBLES_DIB_correlations}) of pairs of DIBs, and a number of families of DIBs have been proposed (e.g., \citealt{Fan_2022_DIB_families_data_driven, Lan_2015_DIB_correlations_SDSS}). Correlation with environment has also been used to define DIB classes, for example by whether a DIB is stronger in $\sigma$-type clouds (lower density, higher UV radiation environments, after $\sigma$ Sco) or $\zeta$-type clouds (higher density, lower UV radiation environments, after $\zeta$ Oph). Up until recently, the optical extinction curve had only been measured with fine wavelength resolution along a few hundred sightlines in the Milky Way (e.g., \citealt{Valencic_2004_UV_extinction_curves, Gordon_2009_FUV_extinction_curves, Fitzpatrick_2019_extinction_curves}), limiting the power of correlation studies between the broad extinction curve and individual extinction features.

Prior to the availability of low-resolution spectra from the European Space Agency's Gaia mission, the largest catalogs of $R(V)$ contained $\sim$37,000 sightlines \citep{Schlafly_2016_Rv_catalog, Schlafly_2017_Rv_3D_mapping}, and were based on broadband photometry. \citet{Lallement_2024_A_A} correlated these $R(V)$ measurements with the DIB catalog based on Gaia Radial Velocity Spectra \citep{Schultheis_2023_Gaia_DIB_catalog}, finding a positive correlation between $R(V)$ and the strength of the \gDIB. However, the crossmatch between these catalogs is relatively small, and the extinction curve catalog only captures a single dimension of variation ($R(V)$), which does not capture the ISS and VBS broad optical extinction-curve variation. Additionally, the DIB catalog used \citep{Schultheis_2023_Gaia_DIB_catalog} was shown to be significantly contaminated by Fe lines \citep{Saydjari_2023_ApJ}.

\begin{figure*}[htb!]
    \centering
    \includegraphics[width=\linewidth]{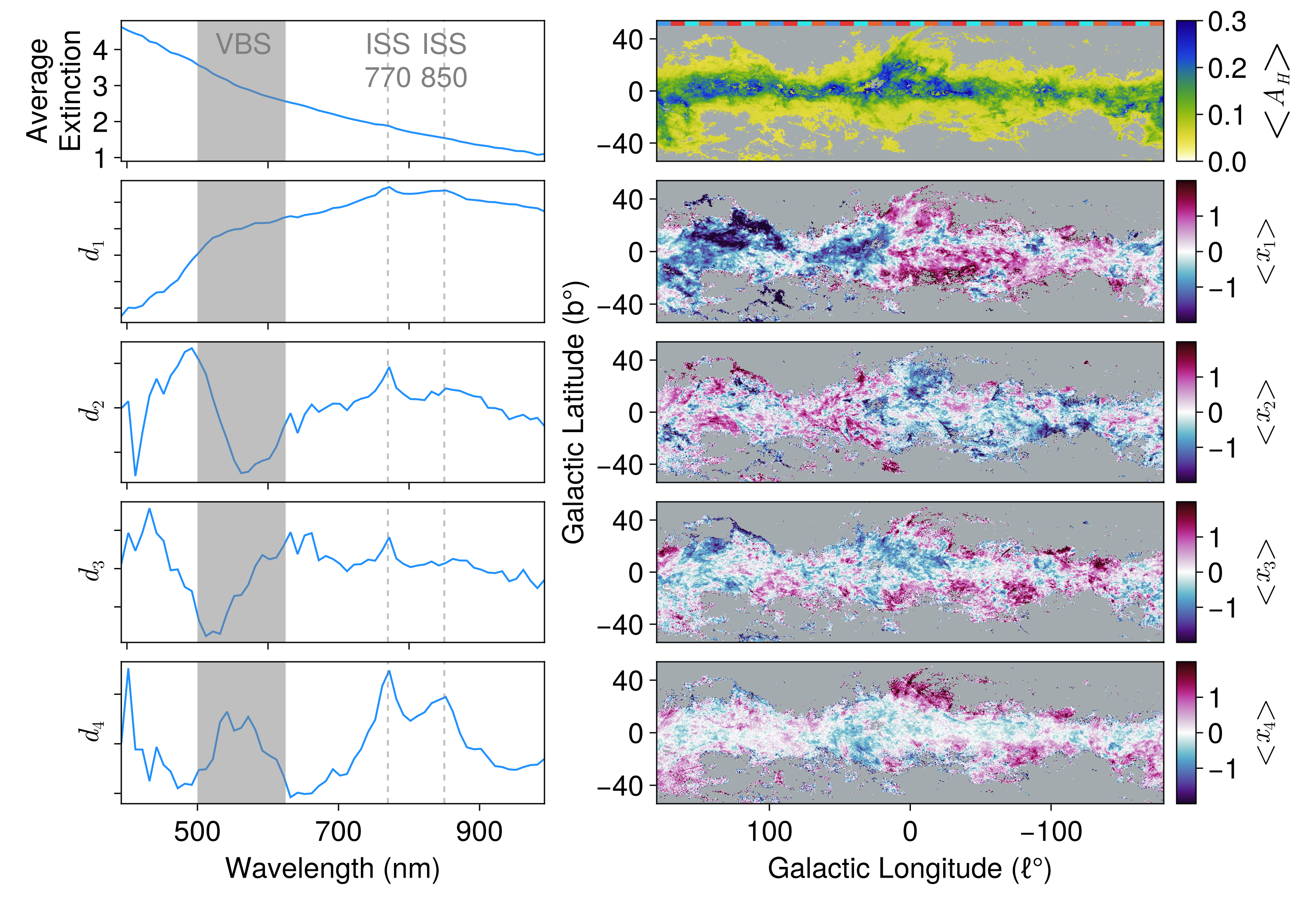}
    \caption{\textbf{Left:} Average extinction curve and first four extinction-curve components of its variation inferred using Gaia XP spectra by \cite{Green_2024_arXiv}. \textbf{Right:} The median H-band extinction and coefficients associated with the extinction-curve components, after the sign changes and standard scaling described in Sections \ref{sec:coefficients} and \ref{sec:cuts}. Pixels with less than 4 stars at HEALPix NSIDE 128 are excluded. Strips 10$\degree$ wide defining subsets for jackknifing and test-train (blues-reds) split are indicated in the extinction plot along the top of the panel.}
    \label{fig:extinction_curve_decomp}
\end{figure*}

Gaia Data Release 3 has dramatically increased the number of available optical stellar spectra, publishing 220~million low-resolution, flux-calibrated, optical ``BP/RP'' (or ``XP'') spectra covering the entire sky \citep{Prusti_2016_Gaia_mission, Vallenari_2023_Gaia_DR3_content, DeAngeli_2023_Gaia_DR3_BPRP_processing, Montegriffo_2023_Gaia_DR3_BPRP_calibration}. \citet{Zhang_2025_Sci} measured dust $R(V)$ for 130~million Gaia XP spectra, and \citet{Green_2024_arXiv} measured higher-order extinction-curve variations for 24 million XP spectra. At the same time, large high-resolution spectroscopic surveys have vastly expanded the number of available DIB measurements. \citet{Saydjari_2023_ApJ} has measured DIB strengths in 7,789 high-resolution Gaia RVS spectra (making significant qualitative improvements to previous measurements using Gaia RVS), while \citet{Saydjari_APOGEE_DIB_unpub} has measured 145,713 DIB strengths from ground-based APOGEE spectra. With these large, overlapping catalogs of optical extinction-curve and DIB measurements, it is now possible to search for correlations between features across a range of wavelength scales.

In this paper, we use simple linear models to show that some of the scatter in the DIB-extinction relation is correlated with broad optical extinction features, and therefore likely has a physical origin. Most of the DIBs increase in strength with increasing $R(V)$, though we identify one atypical DIB (at 15616~\r{A}) that behaves in the opposite manner. We additionally calculate high-resolution, near-infrared (H-band) ``spectral response functions'' that trace the change in DIB lineshape as the broad shape of the optical extinction curve changes. These spectral response functions show not only how DIB equivalent width changes with $R(V)$ and higher-order, broad extinction-curve features, but also how DIB profiles change. We find evidence that different broad extinction features are correlated with distinct modes of chemical variation.

In Section~\ref{sec:data} we introduce the extinction (\ref{sec:coefficients}) and DIB datasets (\ref{sec:MADGICS}), cuts (\ref{sec:cuts}), and methods (\ref{sec:MADGICS} and \ref{sec:linfit}) used in this work. In Section~\ref{sec:15272}, we explore the DIB-extinction relation for the \apDIB in APOGEE, with both conventional modeling (\ref{sec:15272_ew}) and using an extended model that includes correlations with $R(V)$ and other extinction features (\ref{sec:15272_seq}). We also provide validation and interpretation for our extended model (\ref{sec:validInterp}) and apply the extended model to each wavelength bin of the spectral residuals (\ref{sec:15272_spec}). We repeat this analysis for the \gDIB in Gaia RVS spectra in Section~\ref{sec:8623}. We provide reproduction code and data in Section~\ref{sec:dataavil} and conclude in Section~\ref{sec:conc}. Further validations and views are provided in the Appendices.

\clearpage

\section{Data and Methods} \label{sec:data}

We compare two types of measurements for each sightline:
\begin{enumerate}
  \item Measurements of extinction and the broad shape of the extinction curve, based on Gaia~XP spectra (and 2MASS/WISE photometry), \S \ref{sec:coefficients}.
  \item Measurements of the DIB spectrum in the NIR, based on APOGEE and Gaia~RVS spectra (both the DIB equivalent width and full spectral residuals), \S \ref{sec:MADGICS}.
\end{enumerate}

We use linear fits with uncertainties in both the dependent and independent variables to study their relationship, \S \ref{sec:linfit}.

\subsection{Extinction Curve Coefficients} \label{sec:coefficients}

We use an extinction-curve decomposition of 24 million stars with Gaia XP spectra \citep{Green_2024_arXiv}. These decompositions are based on empirical extinction curves, obtained by comparing observed Gaia XP spectra to extinction-free model spectra from \cite{Zhang_2025_Sci}. These extinction-free spectra were determined by fitting the Gaia~XP spectra, augmented with parallaxes and near-infrared photometry, using a data-driven forward model that includes intrinsic stellar parameters (effective temperature, metallicity and surface gravity), stellar distance, and foreground extinction \citep{Zhang_2025_Sci}. This model treats extinction as a single-parameter family of curves, which is similar to the conventional parameterization using $R(V)$.

The data-driven model in \cite{Zhang_2025_Sci} works best for FGK stars on the main sequence and giant branch, but struggles with very cool or hot stars, sub-dwarfs and white dwarfs. These problematic sources are removed using quality-of-fit measures, such as reduced $\chi^2$. The model outputs a predicted best-fit extinction-free spectrum for each star, which can then be compared to the observed Gaia~XP spectrum to obtain an empirical extinction curve. \cite{Green_2024_arXiv} used a Bayesian variant of principal component analysis to compute the 16 dominant vector components of the highest-quality $\sim$30,000 extinction curves obtained through that comparison. \cite{Green_2024_arXiv} then projected 24 million extinction curves onto these dominant vector components.

We use only the first four extinction-curve components, which are least affected by any obvious systematics associated with the Gaia observational pattern (``scanning pattern'') and have been shown to be the most physical (see Section 4.1 and Appendix D in \citealt{Green_2024_arXiv}). We modify those components, which have an arbitrary sign, such that all of the correlations of their coefficients with \apDIB residuals in Section~\ref{sec:15272} are positive, the implications of which we discuss therein. Thus, the extinction-curve components used in this work $(d_1, d_2, d_3, d_4)$ correspond to $(g_0, -g_1, g_2, -g_3)$ from \cite{Green_2024_arXiv}. The extinction curve components and the average value of their coefficients on the plane of the sky are shown in Figure \ref{fig:extinction_curve_decomp}. These extinction-curve components should be viewed as ``intensive'' quantities, parameterizing changes in the extinction curve for different lines of sight, which are then multiplied by the total ``amount of extinction'' to obtain the observed wavelength dependent extinction for a given source.

\citet{Zhang_2025_Sci} modeled the extinction curve based on Gaia~XP spectra, combined with NIR photometry from 2MASS and WISE and Gaia parallaxes. We also use that paper's measurements of monochromatic extinction at 8620\,\r{A} (which we term $A_{RVS}$) and the extinction in the 2MASS H-band, which \citet{Zhang_2025_Sci} treated as independent of stellar type. Both $A_{RVS}$ and $A_H$ depend on $R(V)$, as measured by \citet{Zhang_2025_Sci}. $A_{RVS}$ is at roughly the same wavelength as Gaia RVS, while $A_H$ covers roughly the same wavelength range as APOGEE. We add 1\% of the average uncertainties in quadrature to the reported uncertainties for $A_H$ and $A_{RVS}$. 

\begin{figure}[b!]
    \centering
    \includegraphics[width=\linewidth]{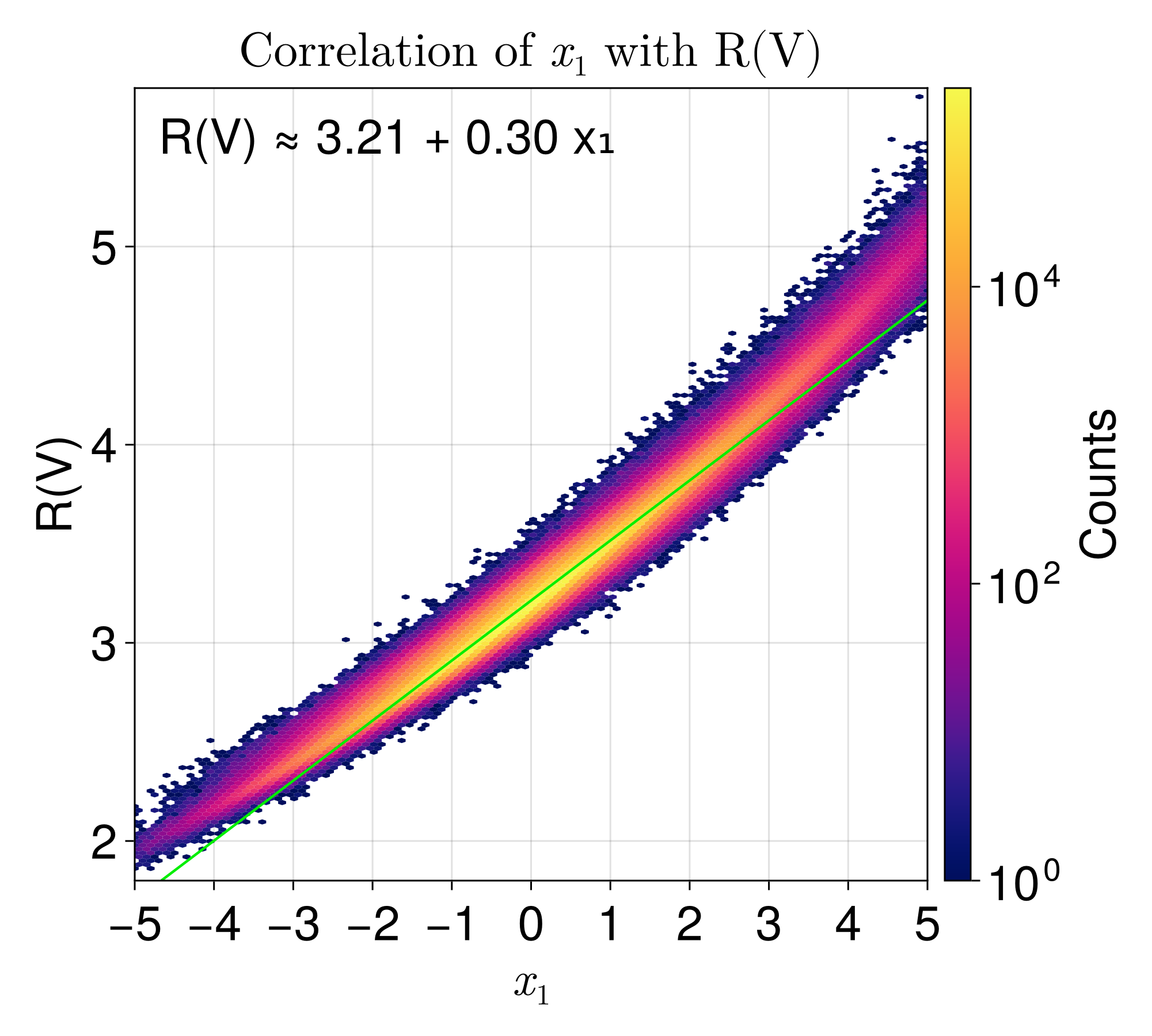}
    \caption{2D histogram of inferred $R(V)$ and $x_1$ extinction curve parameters for the $\pm5$ sigma-clipped sample. A linear relation is fit as a heuristic for interpreting $x_1$.}
    \label{fig:RVx1}
\end{figure}

One might worry that our choice of which extinction measure to use to normalize the DIB equivalent width might imprint correlations between the DIBs and $R(V)$. To alleviate concerns that this impacts our conclusions, we repeated the analyses in this work using three extinction estimates: $A_H$, $A_{RVS}$, and an estimate of $A_V$ from the the Rayleigh-Jeans Color Excess (RJCE) method \citep{Majewski_2011_ApJ}, which we define more precisely in Section~\ref{sec:15272_ew}. All of the key claims of this work are independent of the choice of extinction measure.

\subsection{MADGICS Decompositions} \label{sec:MADGICS}

The spectral datasets are the result of Bayesian component separation pipelines using a method called ``MADGICS,'' Marginalized Analytic Data space Gaussian Inference for Component Separation. In this work, we will focus on the applications of this method to Gaia RVS \citep{Saydjari_2023_ApJ} and APOGEE\footnote{\url{https://github.com/andrew-saydjari/apMADGICS.jl}} \citep{Saydjari_APOGEE_DIB_unpub}, though the method has also been applied in the optical to DESI spectra \citep{Uzsoy_LAE}.

The basic idea of MADGICS is to model a spectrum as a linear combination of components $x_k$, where we put a prior on each of these components expressed as a pixel-pixel covariance matrix $C_k$. That is the components and their priors live in the ``data space'' of spectral wavelength bins. Then, subject to the constraint that the components sum exactly to the data, we obtain a posterior for the components.

\begin{ceqn}
\begin{align} \label{eq:post_mean}
    \hat{x}_{k} &= C_k C_{\rm{tot}}^{-1}\left(x_{\rm{tot}} -\mu_{\rm{tot}}\right) + \mu_k \\
    \label{eq:post_kk}
    \hat{C}_{kk} &= (I - C_k C_{\rm{tot}}^{-1})C_k \\
    \label{eq:post_km}
    \hat{C}_{km} &= -C_k C_{\rm{tot}}^{-1} C_m
\end{align}
\end{ceqn}
where $I$ is the identity matrix, $\hat{x}_k$ is the predicted posterior mean component, $\hat{C}_{kk}$ is the predicted pixel-pixel covariance of pixels in component $k$, $\hat{C}_{km}$ is the predicted covariance of a pixel in component $k$ with a pixel in component $m$, $C_{\rm{tot}}$ is the sum of the covariance priors for all components in the model, and $x_{\rm{tot}}$ is the sum of all components, which is by definition equal to the observed data. For more details see \cite{Saydjari_2023_ApJ} and \cite{Saydjari_MADGICS_unpub}.

For APOGEE, the MADGICS model decomposed the spectra into ``sky continuum'' + ``sky emission lines'' + ``star continuum'' + ``star lines'' + ``DIB'' + ``noise'' components. The DIB component was modeled with a Gaussian lineshape, explicitly sampling over a variable central wavelength and width (first and second moments). \cite{Saydjari_APOGEE_DIB_unpub} only model a single DIB at a time, exploring the two strongest DIBs in the APOGEE wavelength range. From these decompositions, we use the equivalent width measurements and measurement uncertainties for the \apDIB, marginalized over the central wavelength and width. In Section~\ref{sec:15272_spec}, we also make use of the stellar continuum-normalized ``residuals'' component (spectrum) at the posterior mean, shifted to the DIB rest frame using the measured \apDIB velocity.

For Gaia RVS, the MADGICS model can be much simpler because there are no components from the Earth's atmosphere and Gaia only provides the pseudo-continuum normalized specttra. Here the MADGICS model decomposed spectra into ``star lines'' + ``DIB'' + ``noise.'' \cite{Saydjari_2023_ApJ} only modeled the one confidently detected DIB in the Gaia~RVS wavelength range at 8623~\r{A} with the DIB component. Similarly to the APOGEE case, we use the 8623~\r{A} equivalent width measurement and measurement uncertainties as well as the ``residuals'' component (spectrum) at the posterior mean, shifted to the DIB rest frame using the measured \gDIB velocity.

\subsection{Cuts} \label{sec:cuts}

We join the extinction and spectral catalogs on \texttt{GaiaSourceID}. We find it necessary to limit these matched catalogs to high-quality sources and to add systematic error adjustments to the reported uncertainties in order to be robust to outliers. A summary of the number of targets passing each of the cuts is shown in Table \ref{tab:cuts}.

\begin{figure*}[htb!]
    \centering
    \includegraphics[width=\linewidth]{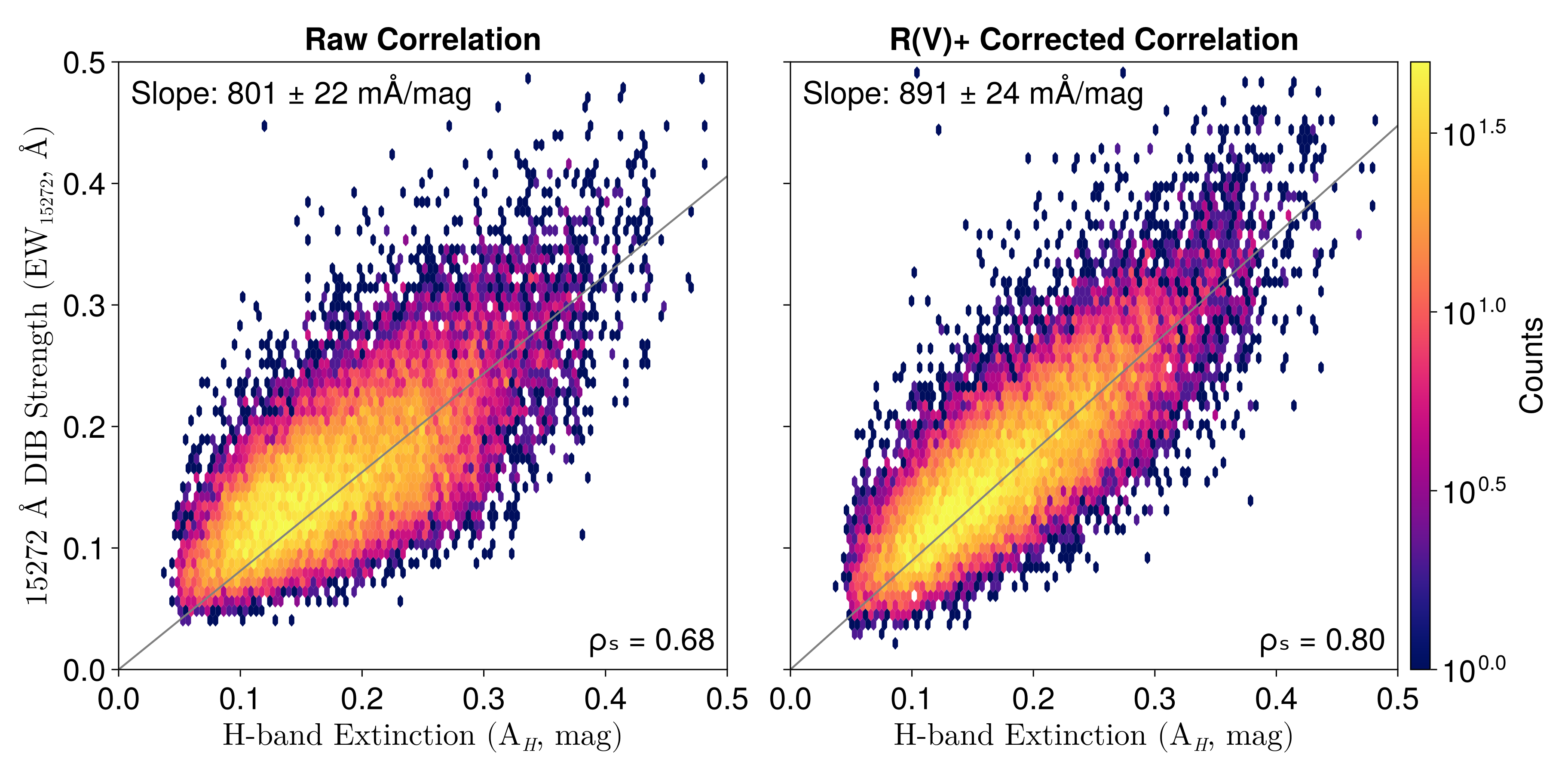}
    \caption{\textbf{Left:} 2D histogram of the DIB strength versus extinction for the \apDIB. Gray line represents best-fit using both extinction and equivalent width uncertainties. \textbf{Right:} The y-axis now shows the DIB equivalent width, after having subtracted off the correlations with extinction curve variations (e.g. $R(V)$, see Equation~\ref{eq:expanded_model}).}
    \label{fig:ext_correlations}
\end{figure*}

\begin{deluxetable}{p{3.5cm}<{\centering}p{2.0cm}<{\centering}p{2.0cm}<{\centering}p{1.1cm}<{\centering}R}[!b]
\tablecaption{Number of Targets in Crossmatch with GaiaXP Catalog also Passing Cumulative Cuts \label{tab:cuts}}
\tablecolumns{3}
\tablehead{Mask & APOGEE & Gaia RVS}
\startdata
DIB Cut & 43,079 & 3,369 \\ 
+ Extinction Cut & 40,327 & 3,236 \\ 
+ $x_k$ Outlier Cut & 40,303 & 3,206 \\ 
\hline
\enddata
\end{deluxetable}

For APOGEE, we restrict the spectral catalog to high-S/N DIB detections (S/N $> 6$), exclude DIBs likely to be blends of multiple velocity components ($\sigma_{\text{DIB}} < 3$~\r{A}), and exclude DIB fits that are not absorptive, in addition to the mandatory quality cuts on $\chi^2$ and grid-edges from \citep{Saydjari_APOGEE_DIB_unpub}. We also add 1\% of the average uncertainty in quadrature to the reported \apDIB EW uncertainty. We added 1\% of the per-wavelength bin average uncertainty in quadrature for the continuum normalized APOGEE residual MADGICS components.

For Gaia RVS, we do the same, cutting at S/N $> 3.8$, $\sigma_{\text{DIB}} < 3$~\r{A}, adding 1\% of the average 8623~\r{A} EW uncertainty in quadrature to the reported values, and 1\% of the per-wavelength bin average uncertainty in quadrature for the continuum normalized RVS residual MADGICS components.

For Gaia~XP, we restrict the extinction catalog to sufficient extinction for stable extinction-curve fits $E > 0.125$ and reasonable reduced-chi-squared values $\chi^2_r < 1.5$. Using these cuts, we normalize the somewhat arbitrary units of the component coefficients by subtracting off their median and dividing by a robust measure of the distribution width (IQR/1.34896, where IQR is the inter-quartile range), such that the data are approximately standard scaled. After this standardization, we call these coefficients the $x_k$ associated with the $d_k$ components. We define a ``high-quality'' sample, masking out points in the heavy tails $-5 < x_k < 5$ ($\sigma$) in any of the $x_k$, $k = 1 - 4$. For the uncertainties on $x_k$, we approximate the $x_k$ as uncorrelated and use the diagonal component of the inverse-covariance matrix in \cite{Green_2024_arXiv}, propagating the above renormalization. We also add 1\% of the average uncertainty for each $x_k$ in quadrature to the reported $x_k$ uncertainty.

As noted in \cite{Green_2024_arXiv}, the coefficient associated with the first component is tightly correlated with $R(V)$, the inverse slope of the optical extinction curve, when this more common single-parameter family of extinction curves is fit to the Gaia XP spectra \citep{Zhang_2025_Sci}. This correlation with our re-normalized $x_1$ is shown in Figure~\ref{fig:RVx1} to help the reader translate correlations with $x_1$ to those with $R(V)$. Over the sigma-clipped range of the ``high-quality'' extinction sample, the relation is almost linear. However, at more extreme values of $x_1$, the relation becomes increasingly non-linear.

\subsection{Linear Fits} \label{sec:linfit}

To help protect against overfitting, we divide the sky into $10\degree$ Galactic longitude slices and do a test-train split that alternates for adjacent strips. This strip width was chosen to be as wide as possible while still providing similar distributions of $A_H$ and the $x_k$ in both the test and train sets. We also use these longitude slices, divided instead into four groups, to help estimate the systematic uncertainty on linear fit coefficients (slopes) via jackknifing. While we have tried to be careful about modeling measurement uncertainties (see below), jackknifing is still extremely useful to provide estimates of systematic uncertainties, for example due to subpopulation variations.

This work is almost exclusively the application of linear models in different contexts to data with sometimes significant uncertainties in both the dependent and independent variables. Thus, linear fits that include uncertainties in both the dependent and independent variables are essential. One method would be Monte Carlo sampling over a full, Bayesian, hierarchal model with priors on the population distribution of independent variables and intrinsic scatter around the linear relation. However, this method is not scalable past a few hundreds of measurements. Another variety of methods go under the name of (weighted) total least squares \citep{markovsky2007overview,Fang_2013_JGeod}, but are usually implemented using matrix factorizations that can even account for correlations between different measurements, but that don't scale well with the number of observations (only tractable for a few thousand). 

In this work, we follow a multidimensional generalization of the method outlined in \cite{Hogg_2010_arXiv} that is tractable on large datasets, as its computational cost only scales linearly with the number of observations. For a given point, we project both the residuals and measurement covariance matrix onto the subspace orthogonal to the proposed linear relationship. Then the log-likelihood, up to a constant independent of the parameters of interest, is just given by Equation~\ref{eq:lin_fit} and we find the maximum likelihood solution by gradient descent \citep[e.g. using LBFGS, ][]{liu1989limited}.

\begin{ceqn}
\begin{align} \label{eq:lin_fit}
    -2\ln(\mathscr{L}) &\propto \sum_{i = 1}^{N_{\text{obs}}} \left[\frac{(\mathbf{\hat{v}}^T\mathbf{Z_i} - \mathbf{\hat{v}}^T\mathbf{b})^2}{\mathbf{\hat{v}}^T\mathbf{S_i}\mathbf{\hat{v}}}\right]\\ \nonumber
    \mathbf{Z} &= \begin{bmatrix} y & x_1 & \cdots & x_k \end{bmatrix} \\ \nonumber
    \mathbf{b} &= \begin{bmatrix} b & 0 & \cdots & 0 \end{bmatrix} \\ \nonumber
    \mathbf{\hat{v}} &= \frac{1}{\sqrt{1 + m_1^2 + \cdots + m_K^2}} \begin{bmatrix} 1 & -m_1 & \cdots & -m_K \end{bmatrix}  \nonumber
\end{align}
\end{ceqn}
This can be thought of as the usual $\chi^2$, but relative to the surface normal to $\mathbf{\hat{v}}$, translated by $\mathbf{b}$. The denominator corresponds to the measurement uncertainty perpendicular to the surface. 

The $\mathbf{Z_i}$ contain one of the $N_{\text{obs}}$ observational measurements for the dependent variable $y$ and $K$ independent variables $x$. In this work, $\mathbf{Z_i}$ is 5-dimensional ($K = 4$), for example in Section \ref{sec:15272_ew} the first entry in is the \apDIB equivalent width and the other four entries are the $x_k A_H$ values for a given star. Since we are interested in how DIB EW, which is an extensive quantity that scales with extinction, correlates with the $x_k$ that are intensive quantities characterizing the extinction curve, we have to either divide EW or multiply $x_k$ by extinction to facilitate this comparison.

The $m_k$ are the ``slopes'' for each of the dependent variables and $b$ is a y-intercept, though in this work we set $b = 0$ throughout. The $\mathbf{S_i}$ are covariance matrices representing the (possibly correlated) measurement uncertainties for the $i$-th measurement. In this work, we take $\mathbf{S_i}$ to be diagonal, ignoring correlated errors between the dependent variables and independent variable. The $\hat{v}$ is a unit vector orthogonal to the proposed linear relation that we use as a projector for the subspace orthogonal to the proposed linear relation.

\section{\apDIB} \label{sec:15272}

\subsection{Equivalent Width Correlation} \label{sec:15272_ew}

DIBs are classically defined as a spectral residual in a velocity frame independent of the stellar velocity and with a strength (equivalent width) that positively correlates with dust extinction. While this relation is usually linear at low dust densities, the correlation can flatten or even turn over at higher dust densities. This so-called ``skin-effect'' \citep{Herbig_1995_ARAA} has been attributed to a decreasing UV interstellar radiation field toward the interior of dense interstellar dust clouds. However, the \apDIB has been shown to have a linear relation with extinction up to A$_{\text{H}} = 1.6$\,mag \citep{Zasowski_2015_ApJ}.

In the linear regime, it is then natural to measure the slope of this relationship between the DIB strength and extinction. The larger the slope, the higher the DIB signal-to-noise for a given amount of dust extinction, which makes the DIB a better tracer of the interstellar medium. The DIB equivalent width is then conventionally modeled with a simple homogeneous linear equation.

\begin{ceqn}
\begin{align} \label{eq:const_model}
    \text{EW}_{\text{DIB}, i} = c_0 \times A_{H,i}
\end{align}
\end{ceqn}
Here $\text{EW}_{\text{DIB},i}$ is the DIB equivalent width (for the $i$-th observation), $A_{H,i}$ is the H-band extinction, and $c_0$ is the slope. The justification for fitting the linear model without a constant term independent of dust extinction is the assumption that the DIB carriers are only present at detectable densities when dust grains are also present. This assumption comes from the fact that both dust and DIBs are prevalent in similar environmental conditions as well as models that take DIB carriers to be constituents that can combine to form larger dust grains or as being formed as fragments from the destruction of larger grains.

\begin{figure*}[htb!]
    \centering
    \includegraphics[width=\linewidth]{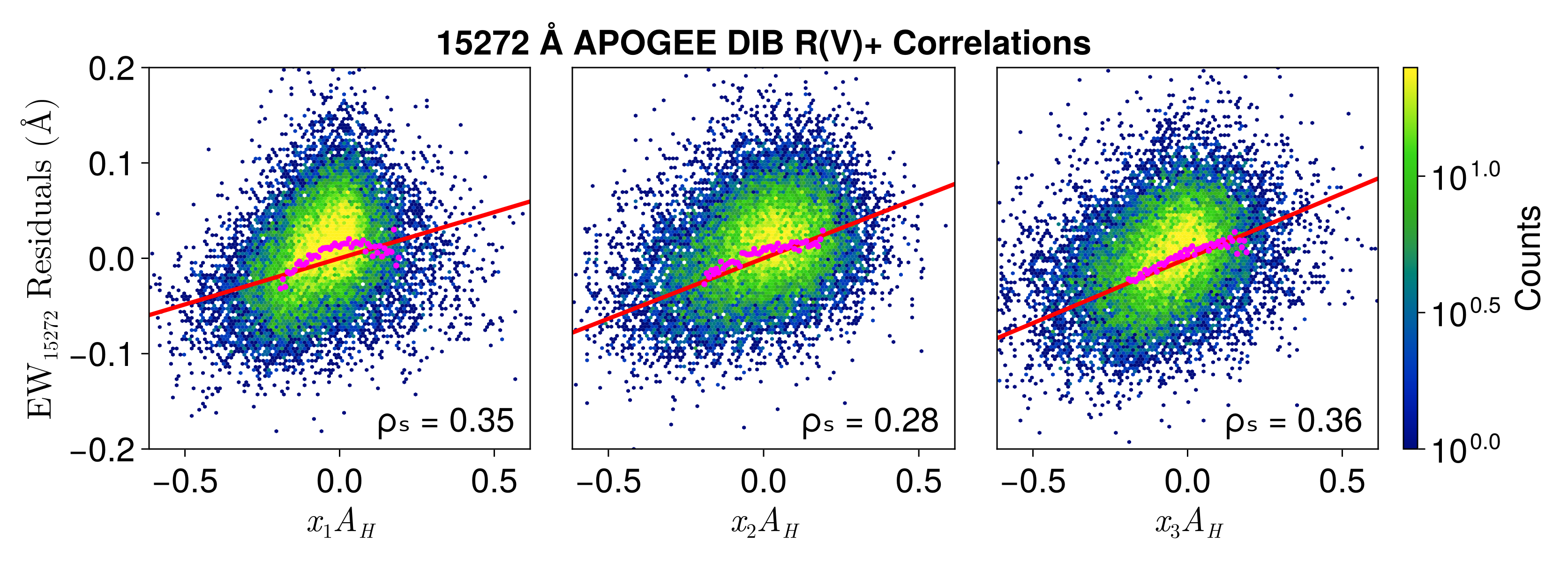}
    \caption{2D Histograms of residuals from $\text{EW}_{\text{DIB}}$ linear extinction fit (Equation~\ref{eq:const_model}) versus products of the derived dust extinction curve coefficients $x_k$ from Gaia~XP spectra with extinction. Joint linear fits per coefficient (Equation~\ref{eq:expanded_model} with only a single $x_k$) using x and y errors are shown as a red line. Magenta scatter points are only a guide to the eye and show the inverse-variance weighted mean $\text{EW}_{\text{DIB}}$ per x-axis bin in the range over which the linear model was fit. The linear fit was \textbf{not} fit to these magenta points.}
    \label{fig:coefficient_correlations}
\end{figure*}

In the left panel of Figure~\ref{fig:ext_correlations}, we use detections of the \apDIB in APOGEE from \cite{Saydjari_APOGEE_DIB_unpub} and extinctions to the stars derived using Gaia~XP spectra from \cite{Green_2024_arXiv} to show this linear relationship. We fit Equation~\ref{eq:const_model} using uncertainties in both the extinction and DIB equivalent width, as described in Section~\ref{sec:cuts}. We find a slope $\text{EW}_{\text{DIB}}/A_H$ of $801 \pm 22$\,m\r{A}\,mag$^{-1}$, where the mean and uncertainty on the slope are estimated by jackknifing as described in Section~\ref{sec:linfit}. 

For comparison, the first large study of the \apDIB by \cite{Zasowski_2015_ApJ} found a slope of $\text{EW}_{\text{DIB}}/A_V = 102 \pm 1$\,m\r{A}\,mag$^{-1}$. In that work, they estimated $A_V$ using the Rayleigh-Jeans Color Excess (RJCE) method \citep{Majewski_2011_ApJ} from WISE and 2MASS photometry, $A_\mathrm{Ks} = 0.918 \times (H - W2 - 0.05)$, and converted to $A_V$ using $A_V/A_\mathrm{Ks} \approx 8.8$ from the $R(V)=3.1$ extinction curve of \citet{Cardelli_1989_Extinction_Curve_CCM89}. When we use this $A_V$ for our extinction variable, we find a slope of $\text{EW}_{\text{DIB}}/A_V = 100 \pm 2$\,m\r{A}\,mag$^{-1}$, which agrees with \cite{Zasowski_2015_ApJ} within the reported uncertainties on both slope estimates. When trying to compare slopes estimated using different extinctions, $\text{EW}_{\text{DIB}}/A_H$ versus $\text{EW}_{\text{DIB}}/A_V$ for example, it is important to account for differences in the extinction curves being assumed. The \citet{Zhang_2025_Sci} family of extinction curves have a mean $A_V/A_H \approx 7.8$, while \citet{Cardelli_1989_Extinction_Curve_CCM89} combined with the $A_H / A_{\mathrm{Ks}} = 1.55$ ratio found by \citet{Indebetouw_2005_IR_extinction} predicts $A_V / A_H \approx 5.68$.

However, a key question remains about the correlation shown in plots such as the left panel of Figure~\ref{fig:ext_correlations}: ``Can any of the scatter around this linear relation be explained by variations in the physical conditions the DIB carriers are experiencing?'' 

\subsection{Higher-Order EW Models} \label{sec:15272_seq}

One tracer we can use for variations in the dust population and the environments they are experiencing is variation in the extinction curve at much larger wavelength scales ($\sim 100$s of nm), such as $R(V)$ which is the inverse of the optical slope of the dust extinction curve. For this purpose, we use the coefficients $x_k$ of the first four components of the extinction curve variation found by \cite{Green_2024_arXiv} for the Gaia XP spectra. In Figure~\ref{fig:coefficient_correlations}, we show the first three of these coefficients and that they have reasonable correlation, Spearman's rank correlation coefficients ($\rho_s$) of $\sim$ 0.3, with the residual scatter around the previously fit linear relationship between $\text{EW}_{\text{DIB}}$ and extinction.

This motivates us to fit a linear model for the DIB strength that depends not only on the extinction, but also the product of extinction and these $x_k$.
\begin{ceqn}
\begin{align} \label{eq:expanded_model}
    \text{EW}_{\text{DIB},i} = c_0 \times A_{H,i} + \sum_{k=1}^{K} c_k \times x_{k,i} A_{H,i}
\end{align}
\end{ceqn}
We fit this model using $K = 4$ of the extinction curve variation coefficients and then subtract off all but the $c_0$ term to show the resulting correlation in the right panel of Figure~\ref{fig:ext_correlations}. By eye, the scatter in the relation between DIB strength and extinction is clearly reduced in the right panel compared to the left panel, which is reflected in the increase in $\rho_s$ from 0.68 to 0.80. This strongly suggests that part of the scatter in the left panel is the result of physical variations in the DIB carrier/dust populations that are well-parameterized by broad (large-wavelength scale) variations in the dust extinction curve. 

Quantitatively, this reduction in scatter can be compared to the error bars on the DIB strength and extinction curve coefficient measurements and corresponds to reducing the width of the distribution of observed $-$ model / observed error-bar (Z-scores) from 1.67 to 1.44~$\sigma$. If the $\text{EW}_{\text{DIB}}$ and $x_k$ uncertainties are well-calibrated, this suggests that while variations with the $x_k$,  $1 \leq k \leq 4$, explain some of the scatter around the linear relation between $\text{EW}_{\text{DIB}}$ and extinction, there is still significant scatter in excess of the measurement uncertainties, with an amplitude of $\sim 44\%$ of those reported uncertainties. 

\begin{figure*}[htb!]
    \centering
    \includegraphics[width=\linewidth]{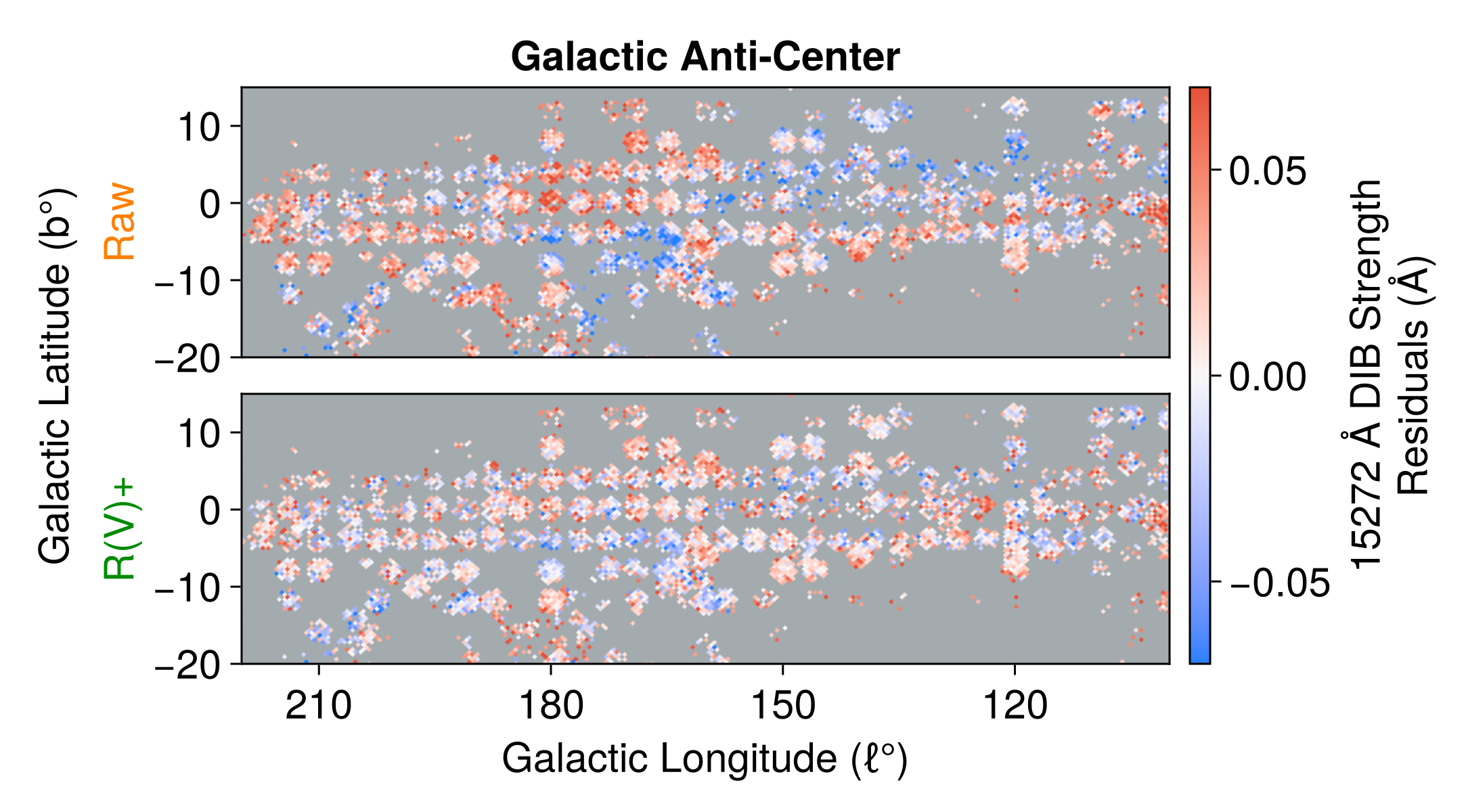}
    \caption{\textbf{Left:} Plane-of-sky median residuals toward the Galactic anti-center for the \apDIB, modeling $\text{EW}_{\text{DIB}}$ as either only a function of extinction (top) or as a function of extinction and extinction variation coefficients (bottom). The extended model reduces the structured residuals.}
    \label{fig:focused_plane_of_sky_residuals}
\end{figure*}

For context, the fractional change in the predicted equivalent width from the linear model in Equation~\ref{eq:const_model} to Equation~\ref{eq:expanded_model} is $17\%$ on average.\footnote{This average is measured by the width of the distribution of fractional changes, since they can have both signs.} If the \apDIB is typical, then this suggests that the variation of DIB strength as a function of extinction curve variation enters at the tens of percent level and is not a small effect!

\subsection{Validation and Interpretation} \label{sec:validInterp}

In Figure~\ref{fig:focused_plane_of_sky_residuals}, we provide one validation of the extension from Equation~\ref{eq:const_model} to Equation~\ref{eq:expanded_model} by showing the plane-of-sky residuals for the $\text{EW}_{\text{DIB}}$ relative to these models. Comparing the top and bottom panels of Figure~\ref{fig:focused_plane_of_sky_residuals} shows that the extended model depending on the extinction curve coefficients $x_k$ significantly reduces the spatially correlated patterns in the plane-of-sky residuals. We also see no evidence of improved performance on the alternating $10\degree$ wide longitude strips used in the test-train split, which suggests we are not in an ``over-fitting'' regime, though we are only fitting 5 numbers to 40,303 data points so ``over-fitting'' is not expected.

\begin{figure}[htb!]
    \centering
    \includegraphics[width=\linewidth]{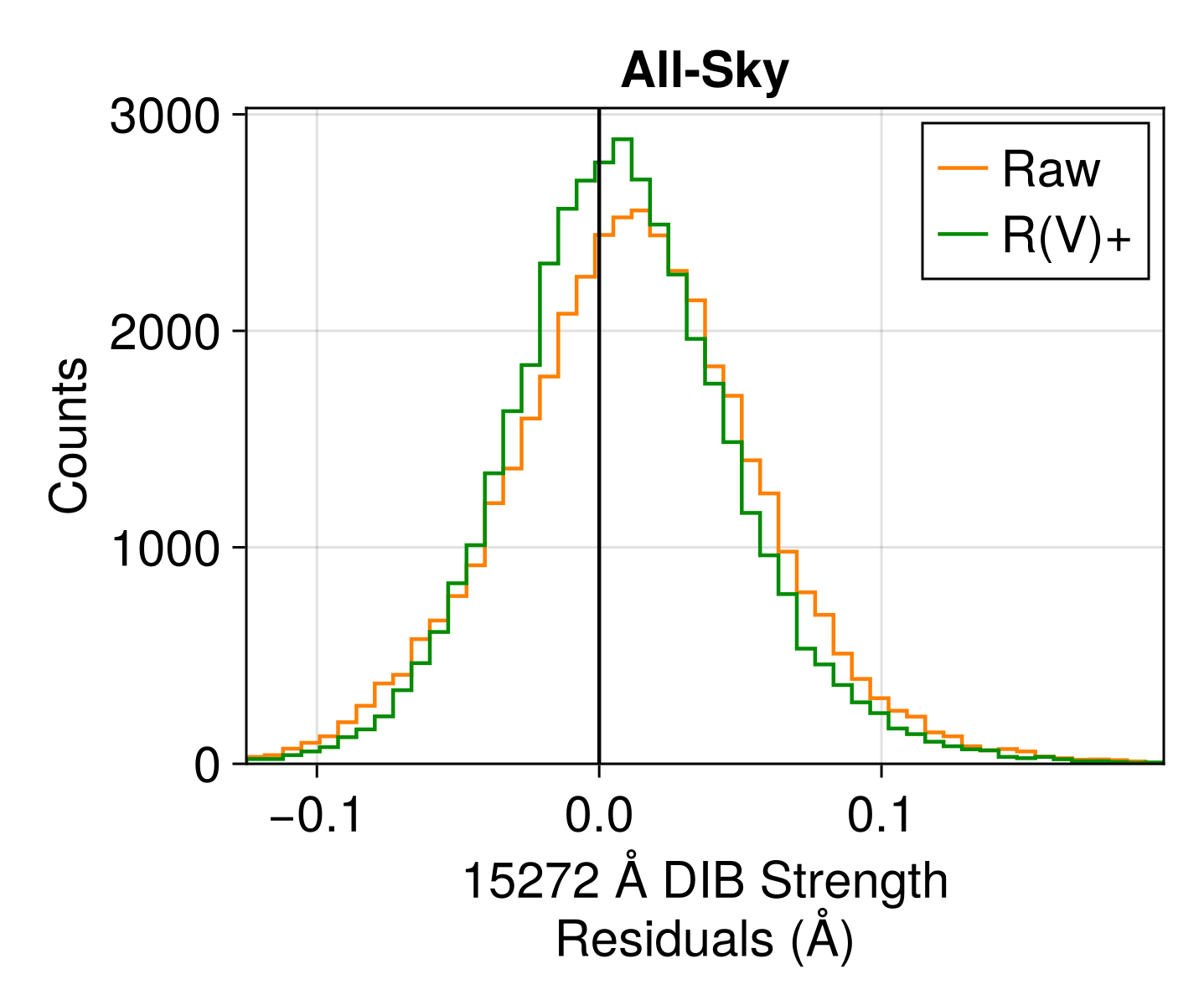}
    \caption{1D histogram of all residuals under extinction only (orange) and extended (green) models.}
    \label{fig:all_sky_res_hist}
\end{figure}

While we picked a specific region to focus on for Figure~\ref{fig:focused_plane_of_sky_residuals}, we show the residuals under both models for the full APOGEE DR17 sky in Appendix~\ref{sec:ap_allSky}. In Figure~\ref{fig:all_sky_res_hist}, we also show a 1D histogram of all of the residuals under both models, Equation~\ref{eq:const_model} in orange, and Equation~\ref{eq:expanded_model} in green. This illustrates that the distribution of residuals are narrower and more centered at zero when including correlations with $R(V)$ and higher order variations in the dust extinction curve. The unweighted median of the residual distribution decreases from 12.2~m\r{A} to 7.2~m\r{A} between the extinction only and extended model. There are still spatially-correlated structured residuals even when using Equation~\ref{eq:expanded_model} in both Figure~\ref{fig:focused_plane_of_sky_residuals} and Appendix \ref{sec:ap_allSky}, which suggests that there are more physical properties driving the scatter in $\text{EW}_{\text{DIB}}$ that remain to be modeled. 

\begin{figure}[t!]
    \centering
    \includegraphics[width=\linewidth]{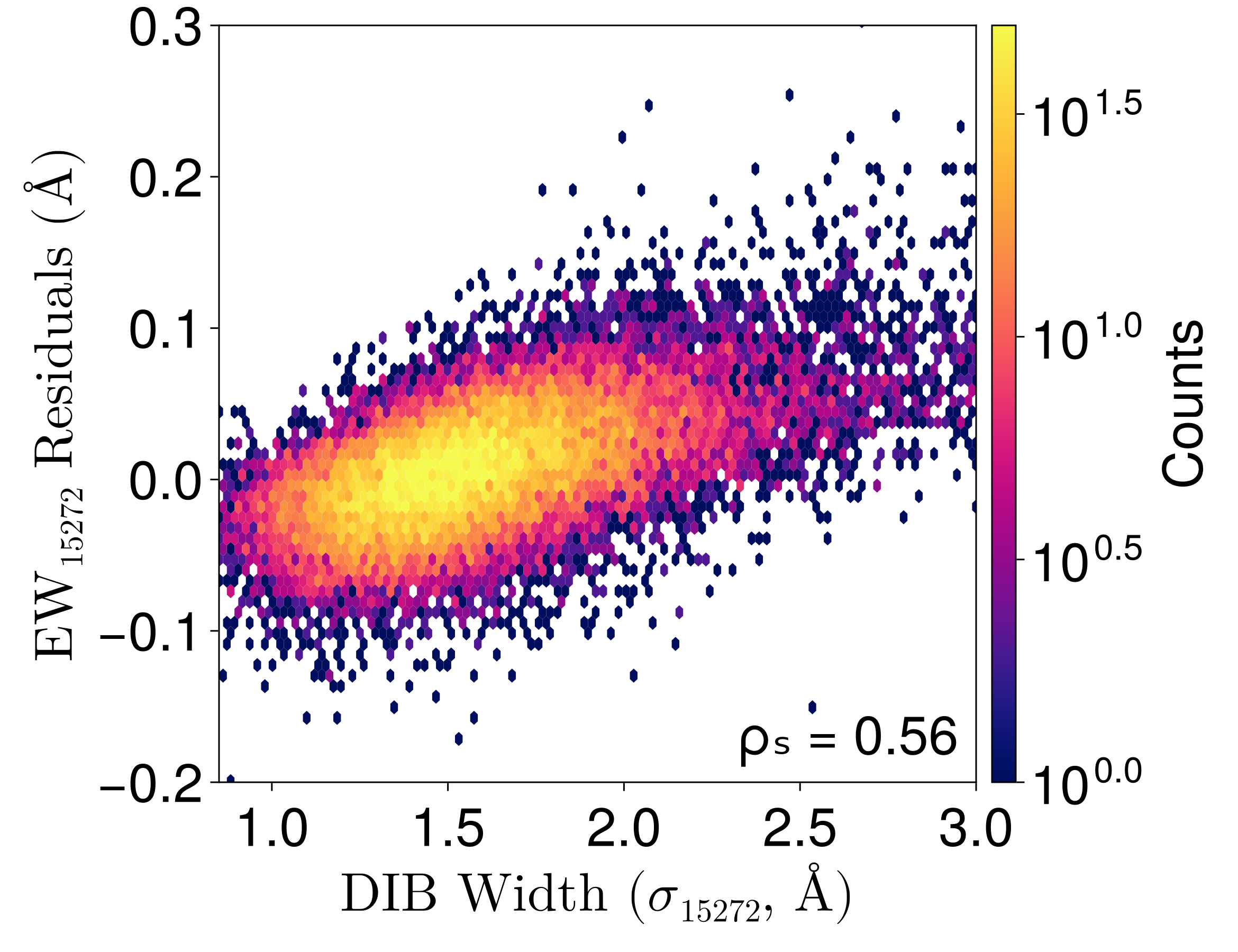}
    \caption{2D histogram of the residuals from EW$_{\text{DIB}}$ fit to extended model in Equation~\ref{eq:expanded_model} versus the Gaussian profile width of the DIB for the \apDIB. The Spearman's rank correlation coefficient between the two is 0.56.}
    \label{fig:dib_sigma_onCorrelation}
\end{figure}

Some of these residuals are strongly correlated ($\rho_s = 0.56)$ with variations in the DIB width, as shown in Figure~\ref{fig:dib_sigma_onCorrelation}. The width of the \apDIB is quite large, $\sim30$~km~s$^{-1}$. For carriers as large as proposed DIB carriers, thermal motions will be negligible\footnote{At these frequencies, thermal broadening for even atomic hydrogen would only be $\sim0.2$~\r{A} at 300\,K, and the effect scales like $\sqrt{T/m}$ with increasing particle mass).} and the typical velocity dispersion from turbulence in the ISM is only $\sim5-10$~km~s$^{-1}$. 

The remaining sources of broadening are lifetime broadening, rotational broadening, and broadening from multiple velocity components along the line-of-sight, each moving at different velocities, where we think the last source is dominant. Thus, to fully explore variations in DIB line broadening due to environmental factors will first require a full 4D reconstruction of the dust density and line-of-sight velocity field to account for broadening due to multiple components along the line-of-sight. We return to the question of rotational broadening in Section~\ref{sec:15272_spec}. However, at this time, it remains an open question which of these mechanisms are dominant in explaining the width variations in Figure~\ref{fig:dib_sigma_onCorrelation} not captured by the extended model in Equation~\ref{eq:expanded_model}.

\begin{figure}[b!]
    \centering
    \includegraphics[width=0.995\linewidth]{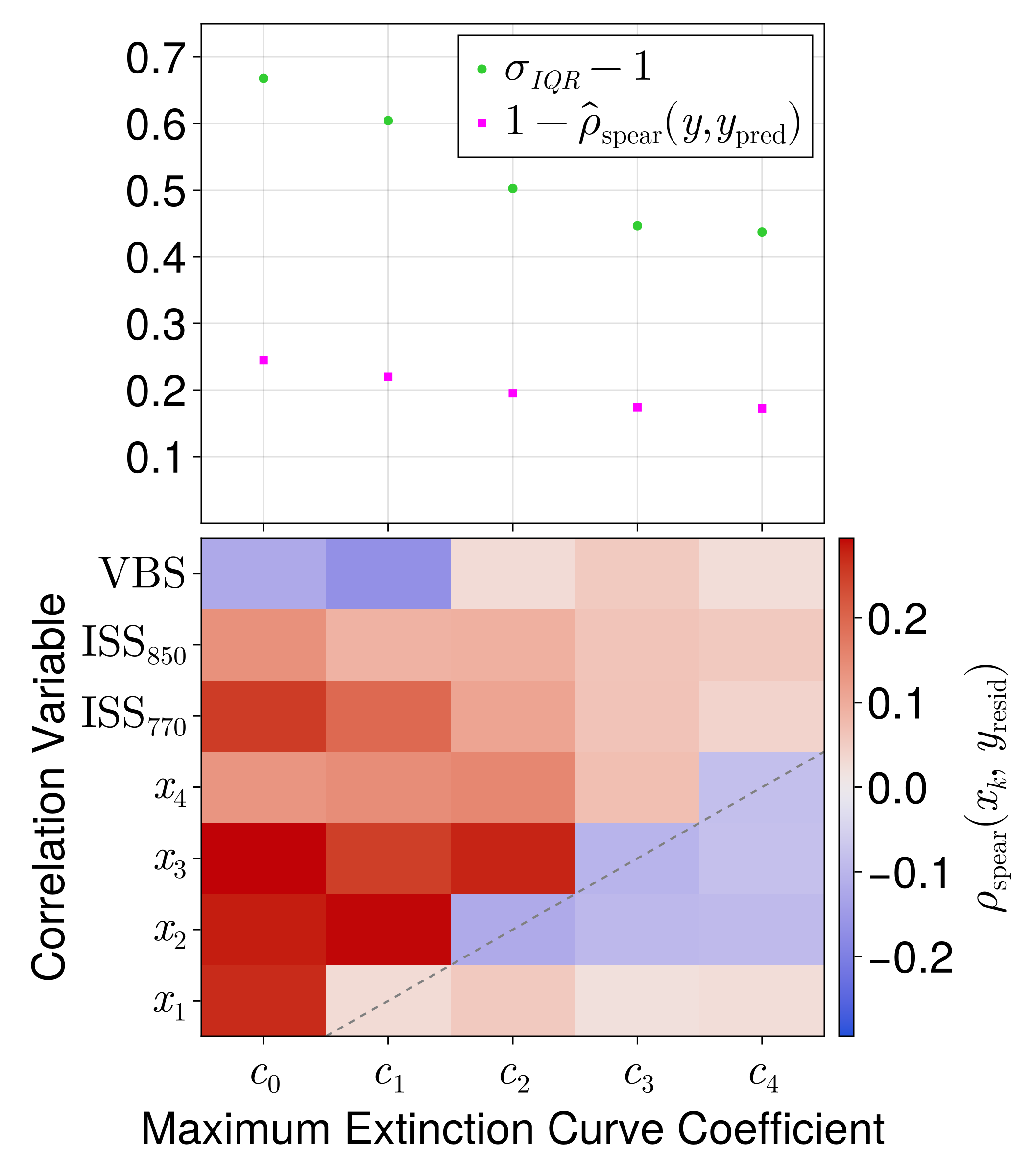}
    \caption{\textbf{Top:} Robust measure of residual scatter relative to reported uncertainties (green, circles) and Spearman's rank correlation coefficient between the model and data normalized for measurement uncertainties (pink, squares) as a function of model complexity (x-axis). \textbf{Bottom:} Spearman's rank correlation coefficient between the residuals and extinction curve coefficients or the equivalent widths of broad wavelength features in the extinction curve.}
    \label{fig:progressive_fits}
\end{figure}

To better understand how to interpret the role of the higher order coefficients, we progressively added each of the $x_k$, fitting models with a progressively higher maximum extinction curve linear coefficient (sequentially adding terms in the sum of Equation~\ref{eq:expanded_model}), the results of which are shown in Figure~\ref{fig:progressive_fits}. In the bottom panel, we show Spearman's rank correlation coefficient between the variables on the y-axis and the observed $\text{EW}_{\text{DIB}}$ as a function of the maximum linear coefficient on the extinction curve parameters, shown on the x-axis. In all cases for Figure~\ref{fig:progressive_fits}, we are showing statistics computed on the ``test-set'' only. After a given $c_k$ has been added to the model, the correlation between the residuals and the corresponding $x_k$ drops to nearly zero, as expected. The fact that the correlation between the residuals and $x_k$ is not exactly zero after a given $c_k$ is included in the model likely reflects the fact that Spearman's rank correlation coefficient does not account for the x and y uncertainties and possible nonlinearity in the relation between the $x_k$ and $\text{EW}_{\text{DIB}}$.

\begin{figure*}[htb!]
    \centering
    \includegraphics[width=\linewidth]{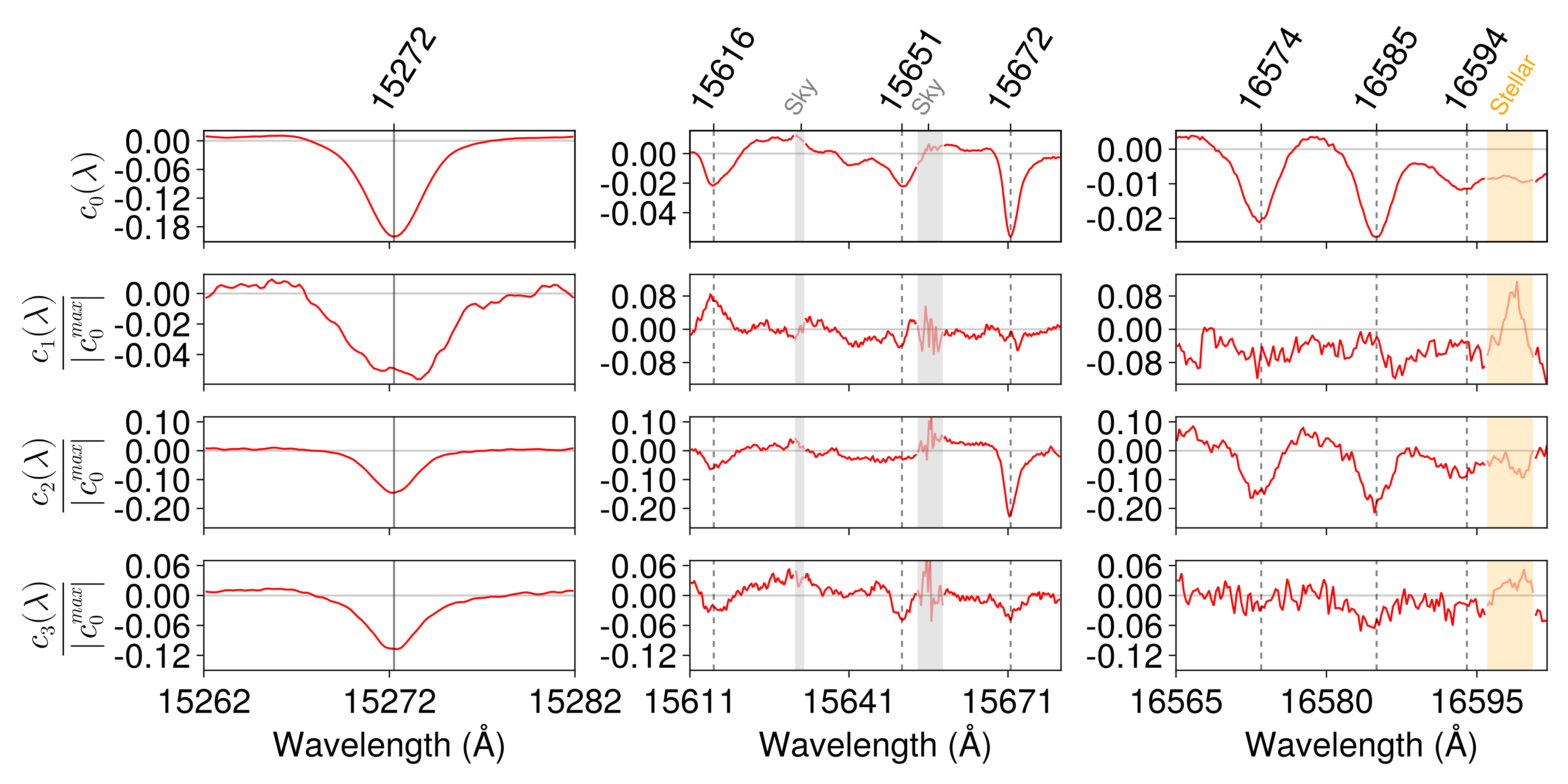}
    \caption{Spectrum showing independent linear coefficients fit to APOGEE spectral residuals in the rest frame of the \apDIB as a function of different extinction curve coefficients $x_k$ (rows) for three wavelength ranges containing DIBs (columns). These ``spectral response functions'' demonstrate that chemical variations accompany variations in the dust extinction curve. See text for more details.}
    \label{fig:linear_response_coefficients}
\end{figure*}

In addition to the $x_k$, we have included the measured equivalent widths of some of the larger-wavelength scale features in the dust extinction curves from \cite{Green_2024_arXiv}. These are the very broad structure, VBS, and both intermediate scale structure features, ISS$_{770}$ and ISS$_{850}$, centered at 7700\,\r{A} and 8500\,\r{A}, respectively. The sign of the correlation of the $\text{EW}_{\text{DIB}}$ residuals with the equivalent widths from these broad structures is positive for both of the ISS features and negative for the VBS. Both of the ISS features have positive EW, indicating an increase in extinction at those wavelengths, while the VBS has negative EW, indicating a decrease in the extinction at those wavelengths. Thus, the sign difference for the VBS is still consistent with increasing amplitude of the (negative) feature with increasing $\text{EW}_{\text{DIB}}$ residuals. \cite{Massa_2020_optical_extinction_structure} suggest the VBS may be simply a lower extinction region between two ISS features at 4870 and 6300~\r{A} that were not explicitly identified and measured by \cite{Green_2024_arXiv}, in which case our observations would be consistent with all four of these ISS features positively correlating with the $\text{EW}_{\text{DIB}}$ residuals.

The amplitude of the broad structure-DIB correlation decreases as we add further extinction curve coefficients. This suggests an interpretation of the $x_k$ in the context of our extended model is that they represent one choice of parameterization of the extinction curve and that together they capture variations in the VBS and ISS equivalent widths. We don't use the equivalent widths of the VBS and ISS directly in our linear expansion of the $\text{EW}_{\text{DIB}}$ because their measurements are noisier than the $x_k$. However, this could suggest alternative dust extinction curve decompositions (using e.g. NMF) for future work.

In the top panel of Figure~\ref{fig:progressive_fits}, we show the evolution of Spearman's rank correlation coefficient $\rho_{\text{s}}$ between the model prediction and data (pink, squares) and a robust measure of the width of the residuals relative to their reported uncertainties (green, circles), both as a function of the model complexity. Ideally both would reach a value of one so the transformed versions we plot would both approach zero from above, however $\rho_{\text{s}}$ is fundamentally limited by the measurement noise, which decreases the correlation. To mitigate that, we actually show $\hat{\rho}_{\text{s}}$, which has been normalized by the correlation between the model values and draws from the model using the noise model of each measurement point.

Clearly, both of these metrics improve (approach unity) with increasing model complexity, before reaching a plateau between $x_3$ and $x_4$, which is part of the motivation for our truncation of the $x_k$ at $k=4$. The smaller amplitude of $\rho_{\text{s}}$ for $x_4$ motivated us to focus on $K=1-3$ in Figure~\ref{fig:coefficient_correlations} and Figure~\ref{fig:linear_response_coefficients}. The gap between these metrics and unity are just another representation of the fact that there remains scatter in the $\text{EW}_{\text{DIB}}$ beyond reported measurement errors that we have not yet explained, even including the linear dependence on variations in the dust extinction curve.

\subsection{ISM Spectrum Correlation} \label{sec:15272_spec}

\begin{figure*}[htb!]
    \centering
    \includegraphics[width=\linewidth]{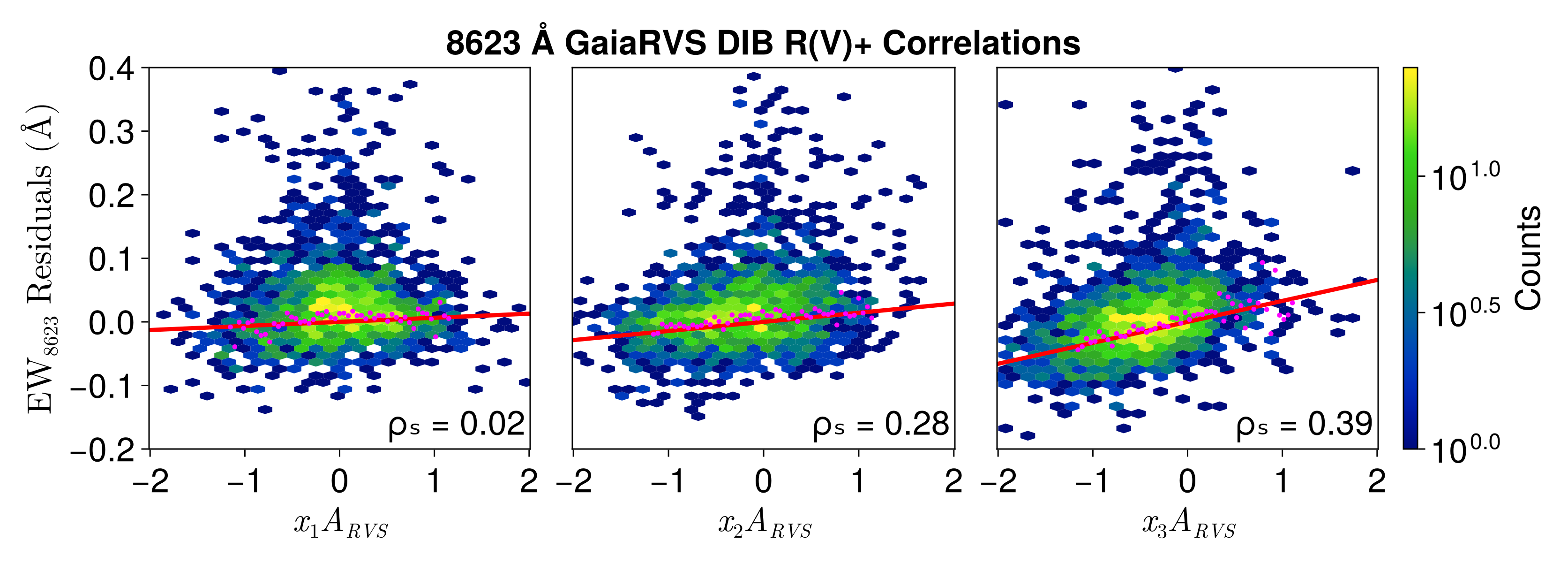}
    \caption{Same as Figure~\ref{fig:coefficient_correlations}, but for the \gDIB measured in Gaia~RVS spectra.}
    \label{fig:coefficient_correlations_8623}
\end{figure*}

In the previous sections, we studied the simplest linear models for a DIB by integrating over the lineshape of the DIB to model only the equivalent width. We can generalize to model each wavelength bin (pixel) in the residual spectrum as a linear combination of the $x_k$ extinction curve coefficients. Computing the $c_k(\lambda)$ basis vectors that minimize the continuum-normalized residuals in Equation~\ref{eq:spec_expanded_model} will give us a better handle on how all of the DIBs and their lineshapes (co-)vary as a function of the $x_k$.
\begin{ceqn}
\begin{align} \label{eq:spec_expanded_model}
    \frac{f_{\text{obs}}-\hat{f}}{\hat{f}_{\text{star continuum}}}(\lambda) = c_0(\lambda)A_H + \sum_{k=1}^{K} c_k(\lambda)x_kA_H
\end{align}
\end{ceqn}
Here $f_{\text{obs}}$ is the observed spectrum. The $\hat{f}$ term is a combination of all of the non-residual model components from a MADGICS decomposition that models APOGEE spectra as a combination of sky continuum, sky emission, star continuum, star absorption lines, and residuals. The $\hat{f}_{\text{star continuum}}$ is just the stellar continuum component of that MADGICS decomposition. Thus, one can think of the left-hand side as the continuum-normalized residuals, using the point estimates for the subtracted/divided components that are the mean of the MADGICS posterior. These basis vectors are defined in the frame of the DIB carriers (i.e. ISM) -- not the stellar frame -- and recall that the $x_k$ are defined so that correlations with the \apDIB feature strength are positive.

Before fitting Equation~\ref{eq:spec_expanded_model}, we have to specify a ``rest-frame'' in which to analyze the spectra. We choose the rest-frame of the strongest DIB in the APOGEE wavelength range, the \apDIB, under the reasonable assumption that different DIB carriers are co-moving. Then Equation~\ref{eq:spec_expanded_model} is basically Equation~\ref{eq:expanded_model}, where the $c_k$ now depend on wavelength and we have applied no regularization to that dependence at all, meaning each wavelength bin is fit entirely independently. Unlike in Sections \ref{sec:15272_ew} and \ref{sec:15272_seq}, we do not perform a test/train split here in order to maximize our signal-to-noise and use all 40,303 spectra passing our cuts (see Section~\ref{sec:data}). We can think of the $c_k(\lambda)$ as DIB spectral response functions (Figure~\ref{fig:linear_response_coefficients}), showing how sensitive the DIB features are to changes in the extinction and extinction curve.

The first column of Figure~\ref{fig:linear_response_coefficients} shows the response of the strongest \apDIB. The strength of the DIB can be read off the first row of this plot, since the magnitude of the $c_0(\lambda)$ indicates the fractional absorption per wavelength bin per mag of H-band extinction. We have chosen the sign here such that negative values indicate absorption, or a decrease in the residuals relative to the stellar continuum. Thus, near the peak wavelength bin, the \apDIB reduces the flux from the stellar continuum by $\sim18\%$ given 1 mag of H-band extinction, though our typical values of $A_H$ range 0 to 0.5. 

In subsequent rows, we divide the $c_k(\lambda)$ by the maximum of $c_0(\lambda)$ for the panel in the top row. This means the subsequent rows can be interpreted as the fractional change in the DIB strengths shown in the top row associated with changing the $x_k$ extinction curve variation coefficients by $1\sigma$ (because of the normalization we performed on the $x_k$ in Section \ref{sec:cuts}). For the \apDIB, this is a 6\% change in the peak with $x_1$ (which is strongly correlated with $R(V)$) and a 15\% and 11\% change with $x_2$ and $x_3$. In subsequent columns, DIBs that have been previously reported in the literature are shown with vertical dashed gray lines near their wavelength centers. Regions in the response spectrum that are noisy due to being dominated by sky emission lines are shaded by a light gray band (2 regions, column 2) and regions dominated by stellar residual contamination are shaded by a light orange band (1 region, column 3).

As mentioned in Section~\ref{sec:data}, we have chosen the arbitrary sign on the components from the decomposition of the Gaia XP extinction curves to be such that the \apDIB strength increases (dip is more negative) as the coefficient on that component increases. We can then compare Figure~\ref{fig:extinction_curve_decomp} to Figure~\ref{fig:linear_response_coefficients} to see the implications this choice has on the broader wavelength features in the extinction spectrum. The main conclusion from that comparison is that directions that increase the strength of the \apDIB, and most other DIBs in this wavelength range, are also the directions that increase the strength of the ISS peaks at 7700\,\r{A} and 8500\,\r{A} (more positive extinction). This is also demonstrated by the sign of the correlation between the $\text{EW}_{\text{DIB}}$ residuals and the equivalent width of the ISS features, shown in the bottom panel of Figure~\ref{fig:progressive_fits}.

There is one notable exception to this trend, which is the 15616~\r{A} DIB. This DIB has a decreasing strength with increasing $x_1$ (which is proportional to $R(V)$). While this DIB has been previously reported in the literature, it is important to be skeptical because not all features in a residual spectrum like $c_0(\lambda)$ are necessarily associated with dust. For example, stellar residuals can be ``broadened out'' by the relative velocity between the intervening dust and the stellar radial velocity and are a common DIB contaminant. To help rule out these cases, we have repeated the fit in Equation~\ref{eq:spec_expanded_model} in both the stellar rest frame and observational rest frame, to search for stellar residuals and telluric residual contaminants, respectively (see Appendix \ref{sec:ap_star_sky_frame}).

As shown in Appendix \ref{sec:ap_star_sky_frame}, while there are some stellar residuals near the 15616~\r{A} DIB, they are small compared to the DIB. In contrast, the feature near 16598~\r{A} (just to the red of the 16594~\r{A} DIB) collapses entirely into a sharp, strong feature in the stellar rest-frame, suggesting that it is predominantly coming from mismodeled stellar lines. We have also performed a series of robustness checks, confirming that the behavior of the 15616~\r{A} DIB (and all of the other features discussed below) are robust to cuts limiting the sample in surface gravity $(\log(g)< 2.2$ , $\log(g)> 2.2)$ or metallicity $([X/H] < -0.2$, $[X/H] > -0.2)$. Thus, we are reasonably confident in the anomalous behavior of the 15616~\r{A} DIB with $x_1$, though it will be important to confirm this with larger sample sizes, such as using the Milky Way Mapper APOGEE targets in SDSS-V.

Beyond the 15616~\r{A} DIB pointing the ``opposite way'' with respect to $x_1$ compared to the other DIBs, the DIBs in this wavelength range also have varying sensitivities to these different $x_k$ extinction curve coefficients. The 15651~\r{A} DIB shows no change in strength with respect to $x_2$, the 16574~\r{A} and 16594~\r{A} DIBs show no change in strength with respect to either $x_1$ or $x_3$ (at least none that is above the noise), and the 16585~\r{A} DIB shows no change in strength with respect to $x_1$ (above the noise). 

Together, the differing behavior of these seven DIBs provides the first observational evidence that there is chemical variation accompanying variations of the broad-band extinction curve. These variations in DIB carrier densities could arise from changes in the interstellar radiation field that change the ionization state of the carriers. The conventional $\sigma$ versus $\zeta$-type DIB classification is based on the former mechanism, where both the \apDIB and \gDIB (see Section \ref{sec:8623}) have been suggested to be $\sigma$-type \citep{MaizApellaniz_2015_MmSAI, Elyajouri_2017_AA}. Comparing the spectral response functions for DIBs of different classes may aid in our understanding of the physical mechanisms driving the broad band variation of the optical extinction curve. Further, we may find that describing DIBs by their correlations with these higher-dimensional extinction curve coefficients brings more nuanced classes and differentiation to describing the behavior of DIBs. Another physical mechanism for the observed chemical variations is compositional changes, for example caused by variations in the C/N ratio in the ISM. While these observations are only correlative, theoretical predictions \citep[e.g. ][]{Zelko_2020} do suggest dust compositional changes could be part of the cause for variations in the broad-band dust extinction curve, in addition to variations in the grain size distribution or PAH fraction.

\begin{figure}[b!]
    \centering
    \includegraphics[width=\linewidth]{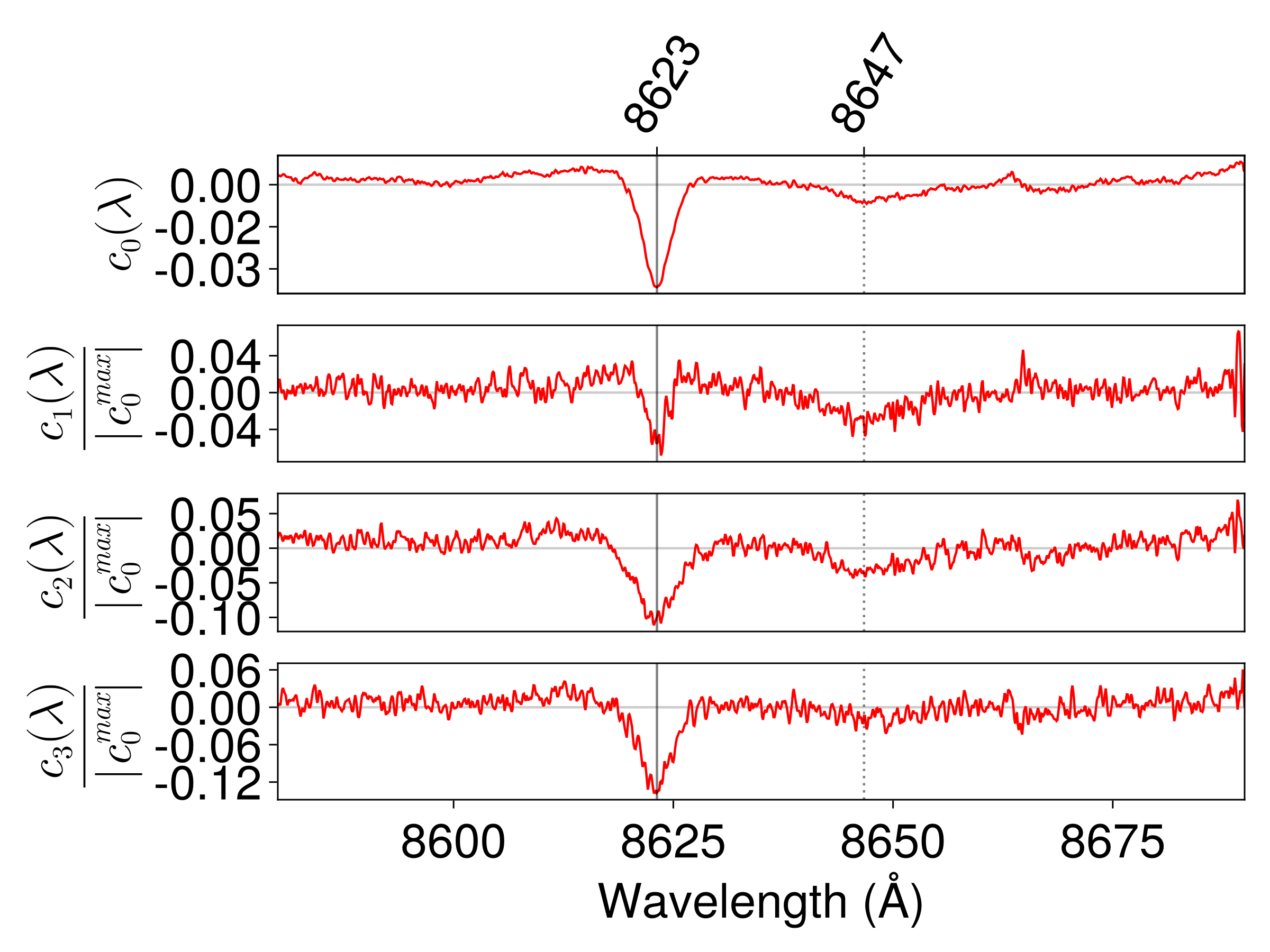}
    \caption{Spectrum showing independent linear coefficients fit to Gaia~RVS spectral residuals in the rest frame of the \gDIB as a function of different extinction curve coefficients $x_k$ (rows). Possible evidence supporting marginal broad 8647~\r{A} DIB detection.}
    \label{fig:linear_response_coefficients_8623}
\end{figure}

The second row, corresponding to $R(V)$ variation, displays interesting variations in the lineshapes beyond the predominantly amplitude variations seen in the DIB response functions with respect to the other $x_k$. Specifically, both the \apDIB and the 15672~\r{A} DIB show a broader, asymmetric, double peaked substructure. The asymmetry in this substructure favors the lower energy (longer wavelength) side. This signature is reminiscent of the P-R branch in rotational transitions, suggesting this broadening could be the result of temperature variations, and thus rotational broadening variations, in directions of increasing $R(V)$. One alternative hypothesis is that these two DIBs have larger velocity dispersion in their populations along these lines of sight, broadening the integrated DIB linewidth for that line-of-sight and causing the DIB response function to resemble a second derivative of the lineshape. A more detailed investigation of this substructure and its origins will be the subject of a forthcoming publication.

\section{\gDIB} \label{sec:8623}

 While the limited number of publicly available RVS spectra in Gaia DR3 restricts the size of the currently accessible \gDIB catalog, we repeated the investigations in Section~\ref{sec:15272}, now using the clean catalog produced by \cite{Saydjari_2023_ApJ} and A$_{\text{RVS}}$. Similar to the \apDIB, we found that the residuals from Equation~\ref{eq:const_model} were correlated with $x_2$ $(\rho_s = 0.28)$ and $x_3$ $(\rho_s = 0.39)$ (Figure~\ref{fig:coefficient_correlations_8623}). Unlike the \apDIB, the correlation is only slight $(\rho_s = 0.02)$ with $x_1$ (which is proportional to $R(V)$). The sign of this correlation is consistent with the correlation between $\text{EW}_{\text{DIB}}/A_V$ and $R(V)^{-1}$ in \cite{Lallement_2024_A_A}. 

Figure~\ref{fig:linear_response_coefficients_8623} shows the spectral response functions (fits to Equation~\ref{eq:spec_expanded_model} with $A_{\text{RVS}}$) for the whole Gaia~RVS wavelength range. The \gDIB increases in strength with increasing ISS strength (positive $x_k$), following the same trend as the \apDIB. The tentatively-claimed, broad DIB at 8647~\r{A} also shows increasing strength with increasing $x_k$ for at least $x_1$ and $x_2$. While this correlation provides further evidence in support of the existence of 8647~\r{A} DIB, it is still difficult to confidently assess the origin of this broad feature. The main difficulty is that the feature lies between the strong \gDIB and the 8662~\r{A} peak of the \CaII triplet and could be the result of the upstream continuum normalization algorithm, which cannot be ruled-out until the Gaia collaboration releases the raw spectra without continuum normalization.

\section{Data/Code Availability} \label{sec:dataavil}
The cross-matched catalogs joining APOGEE and Gaia~RVS spectra to Gaia XP extinctions, Jupyter notebooks, and data products necessary to reproduce all plots in this work are released with this paper and are immediately available on Zenodo at \dataset[doi:10.5281/zenodo.15814994]{https://doi.org/10.5281/zenodo.15814994} (53.3 GB unzipped, 43.5 GB zipped).

\section{Conclusion} \label{sec:conc}

DIB strength does not just depend linearly on extinction; this relation depends meaningfully, at the tens of percent level, on additional parameters describing the overall shape of the optical extinction curve. 

In this work, we combined the three largest catalogs of extinction curve features available. We cross-matched catalogs of DIBs from two different large spectroscopic surveys, APOGEE and Gaia~RVS, with a catalog of dust extinction curves from Gaia~XP spectra and applied simple linear models to study their correlations. In addition to extending classical modeling of DIB equivalent width, we also introduce a powerful new representation of the residuals as ``spectral response functions.`` The main takeaways from this work are below.
\begin{enumerate}
    \item We demonstrate that a significant fraction of the scatter in the DIB strength versus extinction relation is ``real,'' not simply reflecting measurement uncertainties, and reduce the scatter in this relation by expanding the model for DIB strength to depend on additional parameters describing the shape of the optical extinction curve.
    
    \item We demonstrate correlations between dust extinction features across all wavelength scales, suggesting $R(V)$ and the strength of VBS ($\sim 1000$~\r{A} wide), ISS ($\sim 150-500$~\r{A} wide), and DIBs ($\sim 0.4-30$~\r{A} wide) are all influenced by correlated physical mechanisms.
    
    \item We show that only one of the eight DIBs studied in this work, the 15616~\r{A} DIB, appears to decrease in strength along a direction of extinction variation that increases the strength of the \apDIB, and this direction is well-described by increasing $R(V)$.

    \item Two DIBs, the 15272~\r{A} and 15672~\r{A} DIB, show significant broadening and asymmetric lineshapes in their response functions with respect to changes in $R(V)$.

    \item We find the first observational evidence of chemical variation, traced by DIB variation, accompanying $R(V)$ (and other forms of extinction curve) variation.
\end{enumerate}

DIBs trace dust kinematics in 3D, because they bring with them partial distance information in the form of an upper bound on the distance to the backlighting star. Improving our understanding of how DIB strength relates to dust extinction will improve our ability to use measurements of DIBs to reconstruct 3+1D models of ISM kinematics, and will improve extinction estimates in cases where DIBs are used as independent extinction measurements for stars with complicated colors (e.g. embedded YSOs \cite{Carvalho_2022_ApJ}). Further study of line profile substructures that emerged in the spectral response function of the 15272~\r{A} and 15672~\r{A} DIB may help shed light on the physical mechanisms driving $R(V)$ variation across the Galaxy.

\begin{acknowledgments}
We acknowledge helpful discussions with Ruoyi Zhang, Josh Speagle, David W. Hogg, and Karl D. Gordon. Bruce Draine, Eddie Schlafly, and Doug Finkbeiner provided extensive comments on the manuscript and helpful discussions. A.K.S. acknowledges Sophia S\'{a}nchez-Maes for helpful discussions and much support.

A.K.S. acknowledges support for this work was provided by NASA through the NASA Hubble Fellowship grant HST-HF2-51564.001-A awarded by the Space Telescope Science Institute, which is operated by the Association of Universities for Research in Astronomy, Inc., for NASA, under contract NAS5-26555. G.M.G. acknowledges support for this work from the Alexander von Humboldt Foundation, through its Sofja Kovalevskaja Award.

We acknowledge the support by Center for High-Performance Computing staff Brian Haymore, Anita Orendt, and Martin Cuma at the University of Utah.

Funding for the Sloan Digital Sky Survey V has been provided by the Alfred P. Sloan Foundation, the Heising-Simons Foundation, the National Science Foundation, and the Participating Institutions. SDSS acknowledges support and resources from the Center for High-Performance Computing at the University of Utah. SDSS telescopes are located at Apache Point Observatory, funded by the Astrophysical Research Consortium and operated by New Mexico State University, and at Las Campanas Observatory, operated by the Carnegie Institution for Science. The SDSS web site is \url{www.sdss.org}.

SDSS is managed by the Astrophysical Research Consortium for the Participating Institutions of the SDSS Collaboration, including the Carnegie Institution for Science, Chilean National Time Allocation Committee (CNTAC) ratified researchers, Caltech, the Gotham Participation Group, Harvard University, Heidelberg University, The Flatiron Institute, The Johns Hopkins University, L'Ecole polytechnique f\'{e}d\'{e}rale de Lausanne (EPFL), Leibniz-Institut f\"{u}r Astrophysik Potsdam (AIP), Max-Planck-Institut f\"{u}r Astronomie (MPIA Heidelberg), Max-Planck-Institut f\"{u}r Extraterrestrische Physik (MPE), Nanjing University, National Astronomical Observatories of China (NAOC), New Mexico State University, The Ohio State University, Pennsylvania State University, Smithsonian Astrophysical Observatory, Space Telescope Science Institute (STScI), the Stellar Astrophysics Participation Group, Universidad Nacional Aut\'{o}noma de M\'{e}xico, University of Arizona, University of Colorado Boulder, University of Illinois at Urbana-Champaign, University of Toronto, University of Utah, University of Virginia, Yale University, and Yunnan University.

This work presents results from the European Space Agency (ESA) space mission Gaia. Gaia data are being processed by the Gaia Data Processing and Analysis Consortium (DPAC). Funding for the DPAC is provided by national institutions, in particular the institutions participating in the Gaia MultiLateral Agreement (MLA). The Gaia mission website is https://www.cosmos.esa.int/gaia. The Gaia archive website is https://archives.esac.esa.int/gaia.

\end{acknowledgments}

\begin{contribution}

A.K.S. and G.M.G conceived of the project. A.K.S. carried out the analysis, produced the plots, and wrote the majority of the manuscript. G.M.G performed extinction-curve transformations, provided feedback on and suggestions for plots and analysis, wrote the introduction and several paragraphs of the main body, and extensively edited the manuscript. 

\end{contribution}

\facilities{Gaia (XP/RVS), Sloan (APOGEE), Du Pont (APOGEE)}

\software{Julia \citep{bezanson2017julia},
FITSIO.jl \citep{Pence_2010_A_A},
HDF5.jl \citep{hdf5},
Healpix.jl \citep{2021ascl_soft09028T},
Makie.jl \citep{DanischKrumbiegel2021},
Optim.jl \citep{mogensen2018optim},
BenchmarkTools.jl \citep{BenchmarkToolsjl-2016},
Interpolations.jl \citep{Kittisopikul_Interpolations_jl_2025},
FITS \citep{Wells_1981_AAS},
CMasher \citep{2020JOSS52004V},
ColorCET \citep{Kovesi_2015_arXiv},
Jupyter \citep{soton403913}
}

\appendix

\section{All-Sky Residuals} \label{sec:ap_allSky}

\begin{figure*}[htb!]
    \centering
    \includegraphics[width=\linewidth]{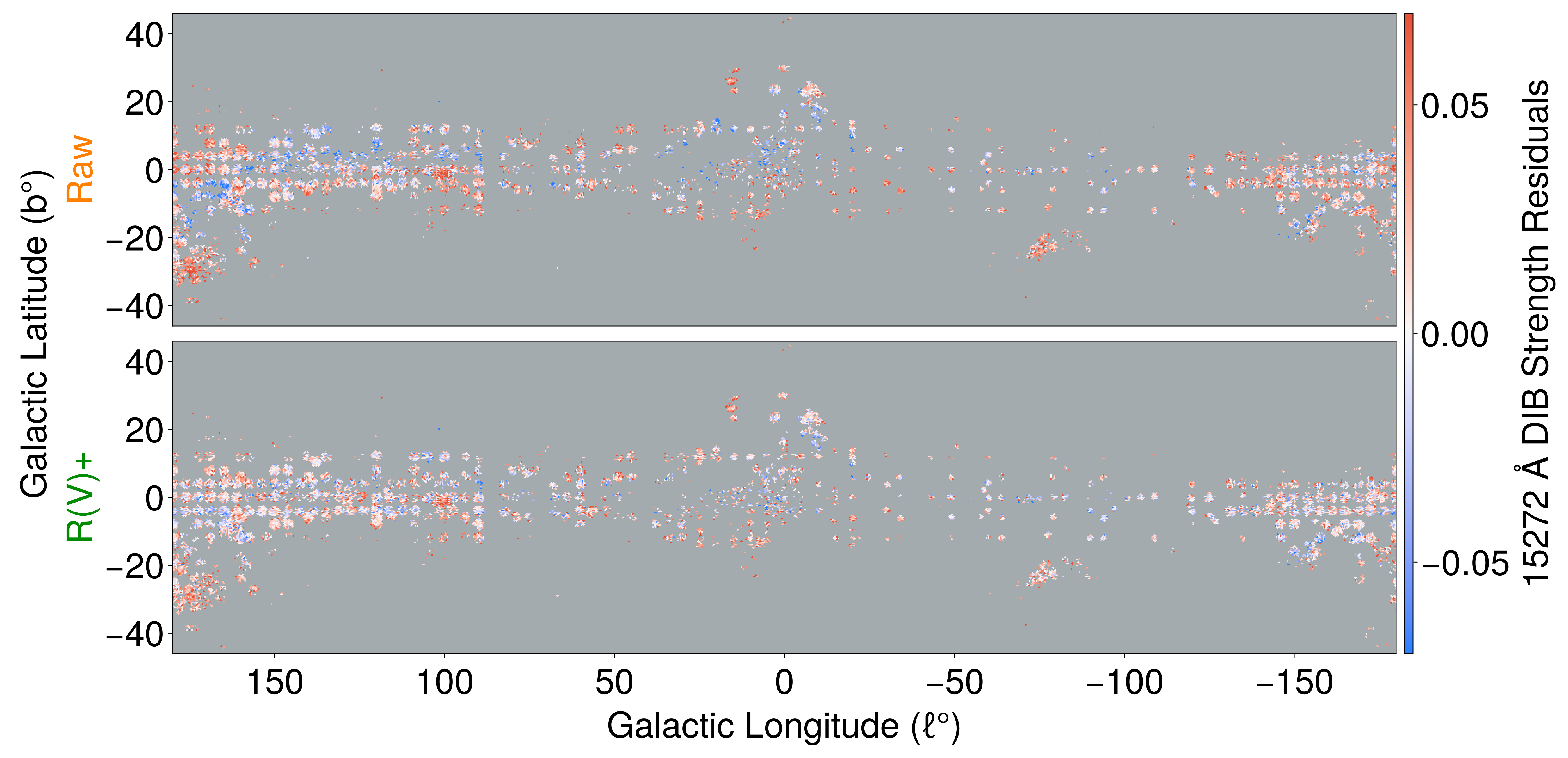}
    \caption{Plane-of-sky median residuals for the \apDIB, modeling $\text{EW}_{\text{DIB}}$ as either only a function of extinction (top) or as a function of extinction and extinction variation coefficients (bottom). The extended model reduces the structured residuals.}
    \label{fig:plane_of_sky_residuals}
\end{figure*}

We present a version of Figure~\ref{fig:focused_plane_of_sky_residuals} in Figure~\ref{fig:plane_of_sky_residuals} that captures the full footprint of DIB detections in APOGEE. Figure~\ref{fig:plane_of_sky_residuals} compares the average plane-of-sky residuals for fit to the conventional model presented in Equation~\ref{eq:const_model} and our extended model in Equation~\ref{eq:expanded_model} for the \apDIB. The extended model significantly reduces the plane-of-sky correlations in the residuals, but clear patterns remain, suggesting that there are more physical properties driving the scatter in $\text{EW}_{\text{DIB}}$ that remain to be modeled. The ``all-sky'' view is presented to help with hypothesis generation on these additional physical origins.

\section{Star- and Sky-frame Residual Fits} \label{sec:ap_star_sky_frame}

\begin{figure*}[htb!]
    \centering
    \includegraphics[width=\linewidth]{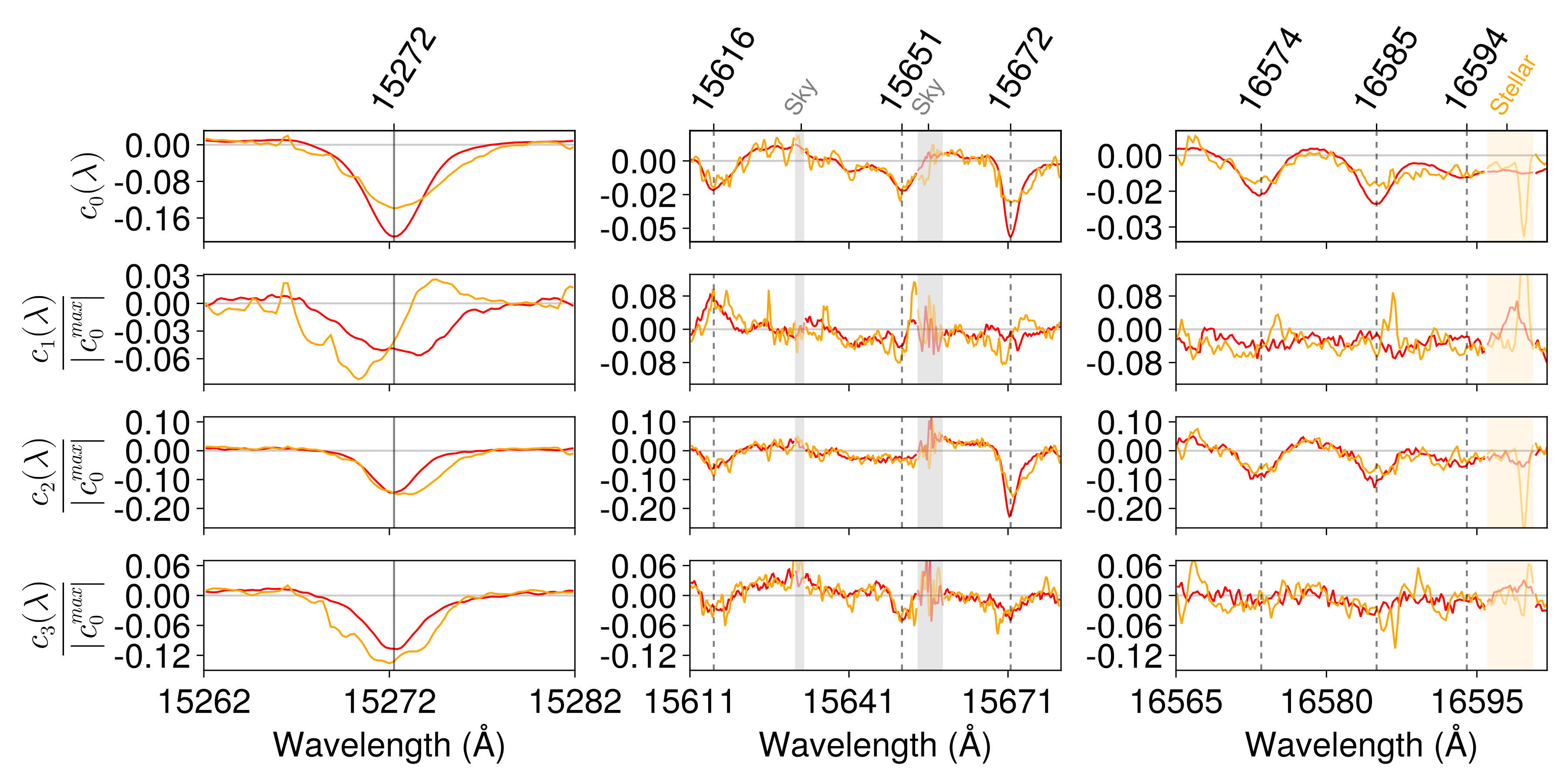}
    \caption{Spectrum showing independent linear coefficients fit to APOGEE spectral residuals in the rest frame of the \apDIB (red) and star (orange) as a function of different extinction curve coefficients $x_k$ (rows) for three wavelength ranges containing DIBs (columns). Features sharper in the star frame than the DIB indicate spurious features likely associated with stellar residuals.}
    \label{fig:starDustFrame}
\end{figure*}

In Figure~\ref{fig:starDustFrame} we show the fits to Equation~\ref{eq:spec_expanded_model} when the spectral residuals are shifted to the ``rest-frame'' of the star. This allows us to identify which features in Figure~\ref{fig:linear_response_coefficients} are likely spurious. Features that are sharp in the star frame are likely mismodeled stellar lines, which then are broadened by the relative velocity between the star and DIB frames. An example of this is the feature near 16598~\r{A}, which appears broad in the DIB frame, but all of that peak collapses into a sharp feature in the stellar rest frame. We find it unlikely that any of the features discussed in the main text primarily originate from the star under this analysis, with similar checks in the sky frame also suggesting none of our features of interest originate from the Earth's atmosphere.

\section{Line Profile Variations} \label{sec:lineprofvary}

\begin{figure*}[htb!]
    \centering
    \includegraphics[width=\linewidth]{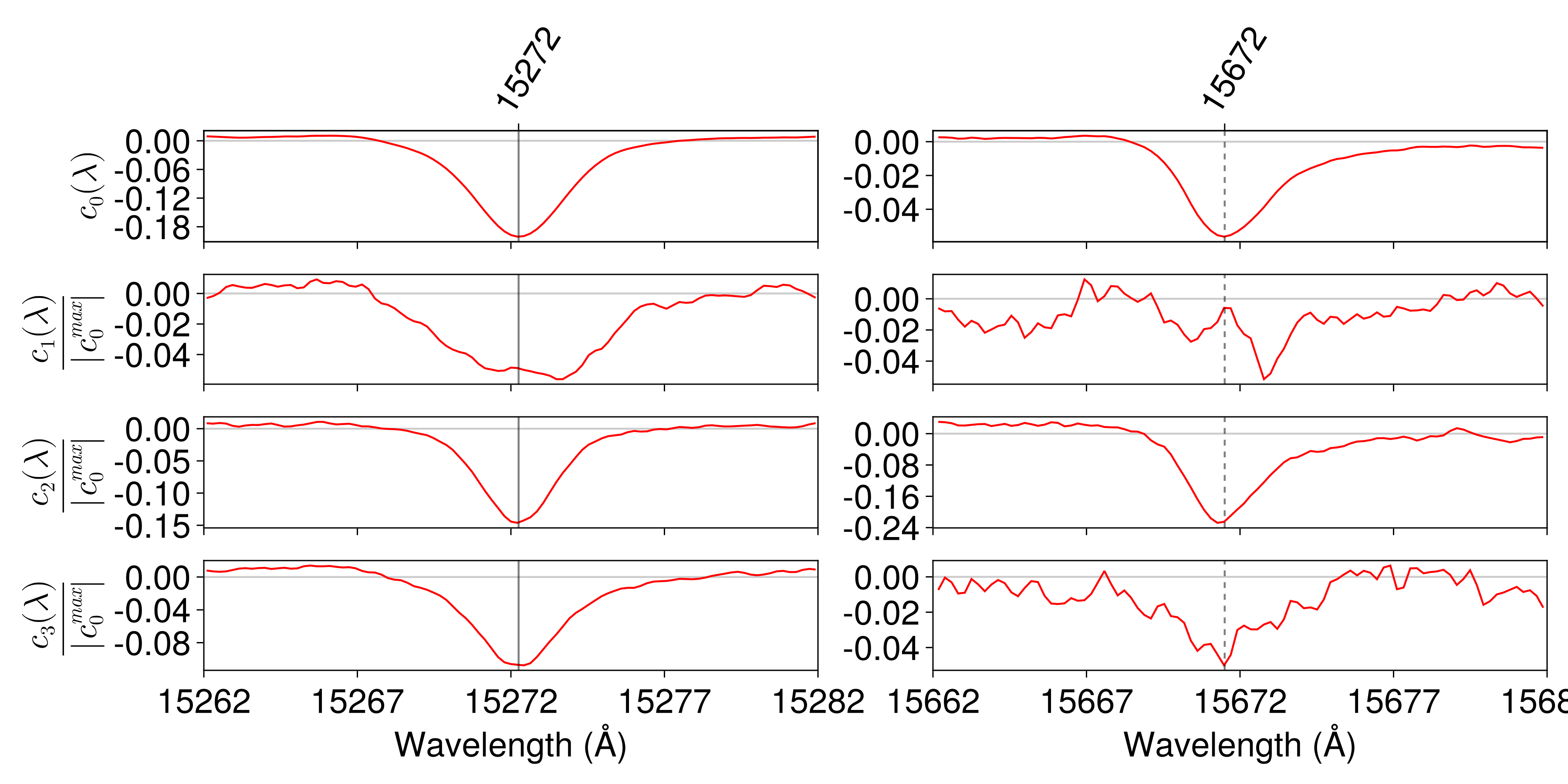}
    \caption{Same as Figure~\ref{fig:linear_response_coefficients}, but focused on the two DIBs, the \apDIB and 15672~\r{A} DIB, that show significant line profile changes in their response to $x_1$ ($R(V)$).}
    \label{fig:splitting_focused}
\end{figure*}

In Figure~\ref{fig:splitting_focused} we show a zoomed in view of the \apDIB and 15672~\r{A} DIB, which both show significant line profile changes in their response to $x_1$ ($R(V)$). 

\clearpage
\bibliography{RVDIBs}{}
\bibliographystyle{aasjournalv7}

\end{document}